\documentclass[aps,pra,twocolumn,superscriptaddress]{revtex4}
\usepackage{amsfonts,amssymb,graphicx}
\usepackage[usenames,dvipsnames]{xcolor}
\usepackage{pdfpages}
\usepackage{footmisc}
\usepackage[normalem]{ulem}
\usepackage{color}

\begin{document}

\title{Exact dynamics of the homogeneous two-qubit $XXZ$ central spin model with the spin bath prepared in superpositions of symmetric Dicke states}
\author{Zejiang Li}
\affiliation{Center for Quantum Technology Research, School of Physics, Beijing Institute of Technology, Beijing 100081, China and Key Laboratory of Advanced Optoelectronic Quantum Architecture and Measurements (MOE), School of Physics, Beijing Institute of Technology, Beijing 100081, China}
\author{Pei Yang}
\affiliation{Center for Quantum Technology Research, School of Physics, Beijing Institute of Technology, Beijing 100081, China and Key Laboratory of Advanced Optoelectronic Quantum Architecture and Measurements (MOE), School of Physics, Beijing Institute of Technology, Beijing 100081, China}
\author{Wen-Long You}
\affiliation{College of Science, Nanjing University of Aeronautics and Astronautics, Nanjing 211106, China}
\affiliation{School of Physical Science and Technology, Soochow University, Suzhou, Jiangsu 215006, China}
\author{Ning Wu}
\email{wun1985@gmail.com}
\affiliation{Center for Quantum Technology Research, School of Physics, Beijing Institute of Technology, Beijing 100081, China and Key Laboratory of Advanced Optoelectronic Quantum Architecture and Measurements (MOE), School of Physics, Beijing Institute of Technology, Beijing 100081, China}

\begin{abstract}
We obtain exact dynamics of a two-qubit central spin model (CSM) consisting of two interacting qubits homogeneously coupled to a spin bath via the $XXZ$-type coupling, with the bath initially prepared in linear superpositions of the symmetric Dicke states. Using the interaction picture Hamiltonian with respect to the non-spin-flipping part of the model, we derive a sequence of equations of motion within each magnetization sector satisfied by the probability amplitudes of the time-evolved state. These equations of motion admit analytical solutions for the single-qubit CSM in which one of the two central qubits decouples from the rest of the system. Based on this, we provide a quantitative interpretation to the observed collapse-revival phenomena in the single-qubit Rabi oscillations when the bath is prepared in the spin coherent state. We then study the disentanglement and coherence dynamics of two initially entangled noninteracting qubits when the two qubits interact with individual baths or with a common bath. For individual baths the coherent dynamics is found to positively correlated to the single-qubit purity dynamics, and entanglement sudden disappearance and revivals are observed in both cases. The entanglement creation of two initially separable qubits coupled to a common bath is also studied and collapse and revival behaviors in the entanglement dynamics are observed. Choosing the equally weighted state and the $W$-class states as the bath initial states, we finally study the dynamics of entanglement between two individual bath spins and demonstrate the entanglement sharing mechanism in such a system.
\end{abstract}

\maketitle

\section{Introduction}
\par The reduced dynamics of simple quantum systems under the influence of structured spin environments~\cite{spinbath2000} has long been a widely studied topic because of its high relevance to quantum decoherence~\cite{Zurek2005,Milburn2005,Sun2006,Petruccione2008,Takahashi2008,Sarma2008,Chen2008,RBLiu2011,F-S2013,Wang2013,CRcat,PRA2014,PRB2016,Cywski2018,Yang2020} and quantum information sciences~\cite{Loss2002,Loss2003,Bose2004,Messina2006,Zhu2007,Cirac2008,Liu2017,Duan2018,Du2019}, excitation energy transfer~\cite{Olaya2008,Sinayskiy2012,Wu2013}, open quantum dynamics~\cite{Petruccione2004,Petruccione2006,Breuer2007,Petruccione2007,Messina2008,Petruccione2012,Petruccione2014}, mathematical physics~\cite{Gaudin1976,Kiss2001,Ortiz2005,Bortz2007,Erbe2010,Claeys2015,Wu2018,Guan2018,Skrypnyk2019,Guan2019,PRB2020,Vill2020}, and also on. Among these, the so-called spin-star network consisting of a preferred central few-spin system coupled to a spin bath without intrabath interactions has attracted much attention due to its exact solvability under certain conditions~\cite{Zurek2005,Milburn2005,Lidar2010,F-S2013,CRcat,Bose2004,Messina2006,Sinayskiy2012,Wu2013,Petruccione2004,Petruccione2006,Breuer2007,Petruccione2007,Messina2008,Petruccione2012,Petruccione2014,Gaudin1976,Kiss2001,Ortiz2005,Bortz2007,Erbe2010,Claeys2015,Wu2018,Guan2018,Skrypnyk2019,Guan2019,PRB2020,Vill2020}. In particular, when the coupling between the central system and the spin bath is homogeneous, the spin bath can be treated as a ``big spin" composed of the surrounding bath spins~\cite{CRcat,Bose2004,Messina2006,Sinayskiy2012,Wu2013,Petruccione2004,Petruccione2006,Breuer2007,Petruccione2012,Petruccione2014,PRA2014,Guan2018,Guan2019}. Although the such obtained homogeneous model looks simple at first sight, it still contains rich physics and possesses elegant mathematical structures. For example, the exact spectrum of a homogeneous isotropic (or $XXX$) central spin model with arbitrary central spin size is determined based on the algebraic properties and integrability of the model~\cite{Guan2018}. It is shown recently that the homogenous anisotropic (or $XX$) CSM with spin-1/2 central spin and arbitrary bath-spin sizes is solvable via the Bethe ansatz~\cite{PRB2020,Vill2020}.
\par In this work, we consider the zero-temperature dynamics of a homogenous two-qubit CSM consisting of two interacting central spins coupled homogeneously to a spin bath via the $XXZ$-type coupling. It is worth mentioning that the dynamics of the two-qubit CSM in its various forms has been studied by utilizing several sophisticated techniques that exploit the ideas and methods of the theory of open quantum systems~\cite{Messina2006,Petruccione2006,Petruccione2012}. In a similar setup, the reduced dynamics of a pair of qubits coupled to individual or common bosonic baths has been widely studied in the context of quantum optics, mainly based on theoretical tools such as Markovian or non-Markovian quantum master equations~\cite{Braun2002,Piani2003,erbly,Kim2006,Science2007,nonmark,Privman2007,Dajka2008,Plastina2008,Aguado2008,Nazir2009,Tanimura2010,Yu2011,Ma2012,Bellomo2013,XBWang2013,Makri2013,Kast2014,Yu2019}. In particular, Yu and Eberly~\cite{erbly} found using Markovian quantum master equations that the initial entanglement between a pair of two-level atoms coupled to their own photonic baths can disappear in a finite time due to the spontaneous emission. Bellomo \emph{et al}.~\cite{nonmark} showed in the same setup that a revival of the disappeared entanglement happens if non-Markovian quantum master equations are adopted. In general, the coupling of central spins to a spin environment can lead to non-Markovian behaviors~\cite{Lidar2007,Breuer2008,Lorenzo2013}, which renders the application of usual Markovian quantum master equations difficult for such systems.
\par Previous studies on the dynamics of central spin problems usually assume thermal states at finite/infinite temperatures for the bath initial state~\cite{Petruccione2008,Cywski2018,Zhu2007,Cirac2008,Sinayskiy2012,Wu2013,Petruccione2004,Petruccione2006,Breuer2007,Petruccione2007,Messina2008, Petruccione2012,Petruccione2014}. In this work, based on the big-spin description of the spin bath, we will focus on one type of bath initial states that can be written as linear superpositions of the symmetric Dicke states~\cite{Dicke}. The permutation symmetry of an $N$-qubit symmetric Dicke state ensures that both the initial state and the time-evolved state are symmetric under the permutation of bath spins, which provides an easy way to analyze the intrabath entanglement dynamics. The linear superposition of the symmetric Dicke states covers several physically interesting cases, e.g., the spin coherent state~\cite{PRA1972}, the $N$-qubit $W$ state~\cite{Wstate},~the macroscopic superposition state~\cite{catstate}, and the fully polarized state~\cite{QB}, etc. Among these, due to the close analogy between the qubit-big spin model and the Jaynes-Cummings model in quantum optics~\cite{CRcat,Guan2019}, the spin coherent state (as an analog of a coherent state for an optical field) turns out to be an interesting option for the bath initial state. It is shown in Ref.~\cite{CRcat} that the collapse and revival phenomena originally observed in the Jaynes-Cummings model~\cite{PRL1990} also occur in the qubit-big spin model if the spin bath is prepared in a spin coherent state. Recently, the central spin dynamics in this system is obtained analytically using a recurrence method based on an expansion of the time-evolution operator~\cite{Guan2019}.
\par The central spin model also offers a suitable platform to study the entanglement sharing among various parts of a composite system. For example, the qubit-big spin entanglement in a simple Ising CSM is shown to be bounded by an entanglement-sharing inequality in terms of the intrabath entanglement~\cite{Milburn2005} for spin baths prepared in $W$-class states. It is thus interesting to investigate the relations between intrabath entanglement and system-bath entanglement or central system decoherence in the spin-flipping $XXZ$ CSM.
\par In this work, we are interested in the reduced dynamics of the two-qubit system under the influence of the spin bath, as well as that of two individual bath spins affected by the central system, both of which can be obtained from the time-evolved state of the whole system. By setting the bath initial state to be a generic linear superposition of the symmetric Dicke states, we obtain the explicit form of the time-evolved state, where the probability amplitudes satisfy a set of equations of motion that can be directly derived by using the interaction picture Hamiltonian with respect to the non-spin-flipping part of the model. In particular, when one of the two qubits decouples from both the other qubit and the spin bath, the equations of motion for the resulting single-qubit CSM have closed-form solutions that help us find a quantitative explanation of the collapse-revival phenomena previously observed in the single-qubit Rabi oscillators~\cite{CRcat,Guan2019}.
\par For a general two-qubit CSM, we focus on the reduced dynamics of the two-qubit system in two situations: (a) The two qubits interact with a common spin bath; (b) Each of the two qubits interacts with its own spin bath, where the two-qubit dynamics is directly determined by the single-qubit dynamics~\cite{nonmark}. With the help of the analytical solution and numerical simulations of the equations of motion, we calculate time evolution of the concurrence~\cite{Wootters} and relative entropy of coherence~\cite{Plenio} for the two-qubit system. In the case of individual baths, we focus on the disentanglement dynamics of two initially entangled qubits without inter-qubit interaction. We find that entanglement sudden disappearance and later revivals happen due to the non-Markovian nature of the spin baths. We also find that the two-qubit coherence dynamics behaves similarly to the single-qubit purity dynamics in the single-qubit problem, indicating that the two quantities are positively correlated. When the two qubits interact with a common bath, we observe bath-mediated entanglement and coherence generation. In particular, for bath initial states with certain polarizations, we find that collapse and revival phenomena can occur in the entanglement and coherence dynamics, which provides a possible scheme to dynamically generate long-lasting steady entanglement or coherence in the collapse regime.
\par The explicit form of the time-evolved state of the whole system also allows us to calculate the reduce dynamics of a pair of individual bath spins. We are particularly interested in the entanglement dynamics of two initially entangled bath spins. Since the spin coherent state is a separable state of all the bath spins, we concentrate on two types of entangled bath states, i.e., an equally weighted superposition of the symmetric Dicke states and the $W$-class states used in Ref.~\cite{Milburn2005} to illustrate the phenomenon of entanglement sharing in the non-spin-flipping Ising CSM. We find that in both the single-qubit and two-qubit CSM the intrabath entanglement is positively correlated to the coherence of the central system, while negatively correlated to the system-bath entanglement. This demonstrate the entanglement sharing mechanism in the general spin-flipping $XXZ$ CSM.
\par The rest of the paper is organized as follows. In Sec.~\ref{SecII} we introduce our model and derive the equations of motion of the probability amplitudes of the time-evolved state using the interaction picture Hamiltonian. In Sec.~\ref{SecIII} we use the results obtained in Sec.~\ref{SecII} to study the reduced dynamics of the single-qubit and two-qubit systems in detail.  In Sec.~\ref{SecIV} we study the relation among the intrabath entanglement, the system-bath entanglement and the coherence dynamics of the central system and demonstrate the entanglement sharing phenomena in our model. Conclusions are drawn in Sec.~\ref{SecV}.

\section{Model and methodology}\label{SecII}
\subsection{The two-qubit homogeneous $XXZ$ central spin model and interaction picture Hamiltonian}
\par
\begin{figure}
\includegraphics[width=.49\textwidth]{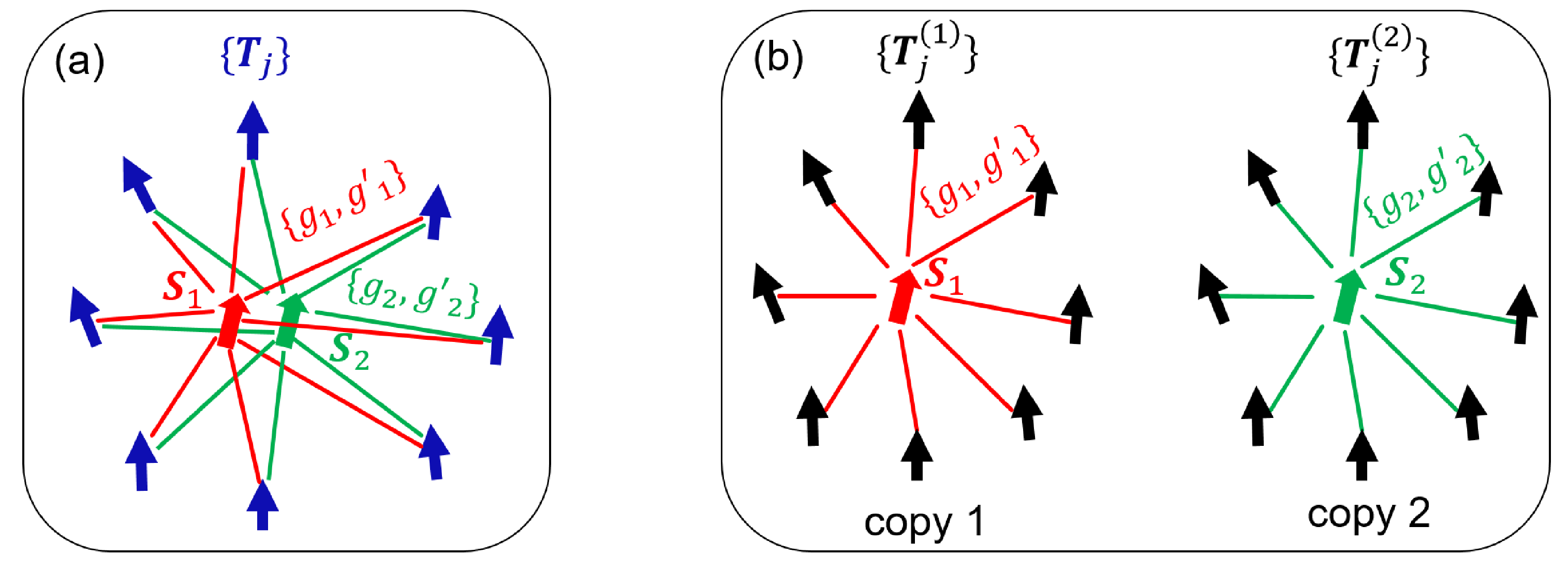}
\caption{(a) Two (interacting) central spins $\vec{S}_1$ and $\vec{S}_2$ interact with a common spin bath via the homogeneous $XXZ$-type coupling [see Eq.(\ref{H})]. (b) Two noninteracting central spins interact with their own spin baths [see Eq.~(\ref{H2q})].}
\label{Fig1}
\end{figure}

 We start with a composite system consisting of two qubits and a spin bath made up of $N$ noninteracting two-level systems [see Fig.~\ref{Fig1}(a)]. Each of the two qubits are assumed to interact homogeneously with the spin bath via the $XXZ$-type coupling (The single-qubit homogenous CSM~\cite{Guan2019} can simply be obtained by turning off the inter-qubit interaction and the coupling between one of the central qubits and the spin bath). Such a system is described by the Hamiltonian
\begin{eqnarray}\label{H}
H&=&H_{S}+H_{SB}.
\end{eqnarray}
Here, $H_S$ is the Hamiltonian for the two qubits that interact with each other via the $XXZ$-type coupling:
\begin{eqnarray}
H_S&=&H_{S,0}+H_{S,1},\nonumber\\
H_{S,0}&=&\omega_1S^z_1+\omega_2S^z_2+2J'S^z_1S^z_2,\nonumber\\
H_{S,1}&=&2J(S^x_1S^x_2+S^y_1S^y_2).
\end{eqnarray}
The system-bath coupling is also assumed to be of $XXZ$-type, and is described by
\begin{eqnarray}
H_{SB}&=&H_{SB,0}+H_{SB,1},\nonumber\\
H_{SB,0}&=&2\sum_{i=1,2}\sum^N_{j=1}g'_i S^z_iT^z_j,\nonumber\\
H_{SB,1}&=& 2\sum_{i=1,2}\sum^N_{j=1}g_i (S^x_iT^x_j+S^y_i T^y_j).
\end{eqnarray}
In the above equations, $S^\alpha_i$ and $T^\alpha_j$ ($\alpha=x,y,z$) are the spin-1/2 operators of the $i$th ($i=1,2$) qubit and the $j$th ($j=1,2,\cdots,N$) bath spin, respectively, $J$ and $J'$ are the qubit-qubit interaction strengths, and $g_i$ and $g'_i$ measure the system-bath coupling strength between the $i$th qubit and any of the bath spins. Note that each qubit interacts uniformly with the spin bath. Since we have assumed that there is no intrabath interaction and the Zeeman energies of the bath spins can be effectively absorbed into the Zeeman term of the central spins according to the magnetization conservation~\cite{PRB2016}, we thus simply omitted the bath Hamiltonian.
\par It is convenient to introduce the collective angular momentum operator for the spin bath,
\begin{eqnarray}
L_\alpha=\sum^N_{j=1}T^\alpha_j,~~\alpha=x,y,z.
\end{eqnarray}
It is easy to verify that the total magnetization of the whole system, $M=S^z_1+S^z_2+L_z$, is conserved. Another obvious conserved quantity is the total angular momentum of the spin bath, $\vec{L}^2=\sum_\alpha L^2_\alpha$. So if we start with an initial state with fixed $l$, where $l(l+1)$ denote the eigenvalue of $\vec{L}^2$, then the evolved state always belongs to this $l$-subspace. We use the simultaneous eigenstates of $\vec{L}^2$ and $L_z$, $\{|l,m\rangle\}$, as the basis states in the $l$-subspace of the spin bath. Note that $\vec{L}^2$ is no longer conserved if either intrabath coupling or inhomogeneity in the system-bath coupling $g_j$ is introduced~\cite{PRA2014}.
\par As will be shown later, to study the real-time dynamics of the system, it turns out to be convenient to work in the interaction picture with respect to the non-spin-flipping part of $H$, i.e., $H_0=H_{S,0}+H_{SB,0}$, where all terms in $H_0$ commute with each other. A straightforward calculation gives the following interaction picture Hamiltonian for $H_1=H_{S,1}+H_{SB,1}$,
\begin{eqnarray}\label{HIt}
&&H_I(t)=e^{iH_0t}H_1 e^{-iH_0t}\nonumber\\
 &=&  J\left[ e^{i 2g'_{12}L_zt}e^{i\omega_{12}t}S^+_1 e^{- i2J' (S^z_1-S^z_2)t}S^-_2+\mathrm{H.c.}\right]\nonumber\\
&+&\left[\sum_{k }g_kS^-_ke^{-i\omega_kt -i2(J' S^z_{\tilde{k}}+g'_kL_z)t}\right]e^{ i\sum_{k }2g'_kS^z_k t}L_+\nonumber\\
&&+\mathrm{H.c.},
\end{eqnarray}
where $S^\pm_k=S^x_k\pm iS^y_k$ ($k=1,2$), $L_\pm=L_x\pm iL_y$, $g'_{12}\equiv g'_1-g'_2,~\omega_{12}\equiv\omega_1-\omega_2$, and $\tilde{1}=2$, $\tilde{2}=1$. The time-dependent interaction picture Hamiltonian $H_I(t)$ can be viewed as an extended two-qubit $XX$ CSM, where the effect of the Ising part of the $XXZ$-interaction in $H$ has been absorbed into the spin lowering/raising operators. The interaction picture Hamiltonian for a single qubit, say qubit 1, can be obtained by setting $J=J'=\omega_2=g_2=g'_2=0$ in $H_I(t)$ given by Eq.~(\ref{HIt}),
\begin{eqnarray}\label{HIt-single}
&&H^{(1)}_I(t)= g_1e^{-i\omega_1t}S^-_1e^{-i2g'_1L_zt}e^{ i2g'_1S^z_1t}L_++\mathrm{H.c.}
\end{eqnarray}
\subsection{Initial states of the spin bath}
\par Throughout this paper, we will focus on the $l=N/2$ subspace of the spin bath, which is spanned by the following $N+1$ symmetric Dicke states~\cite{Dicke}
\begin{eqnarray}\label{Dickestate}
 |m\rangle_D&\equiv& |\frac{N}{2},\frac{N}{2}-m\rangle\nonumber\\
&=&\frac{1}{\sqrt{C^m_N}}\sum_{j_1<j_2<\cdots<j_m}T^-_{j_1}\cdots T^-_{j_{m}}\left(\prod^N_{l=1}|+\rangle_l\right),\nonumber\\
\end{eqnarray}
where $m=0,1,\cdots,N$ counts the number of excitations with respect to the highest weighted state $|0\rangle=|\frac{N}{2},\frac{N}{2}\rangle=\prod^N_{l=1}|+\rangle_l$, $C^n_N=\frac{N!}{n!(N-n)!}$ is the binomial coefficient, $T^\pm_j=T^x_j\pm iT^y_j$ and $|+\rangle_j$ are respectively the spin raising/lowering operator and the spin-up state of the $j$th bath spin. The Dicke state $|m\rangle_D$ is symmetric in the sense that it is invariant under the permutation between any spin pairs in the bath.
\par  We will assume that the initial state of the spin bath is a general linear superposition of the $N+1$ Dicke states,
\begin{eqnarray}\label{phiB}
|\varphi_B\rangle=\sum^N_{m=0}\gamma_m |m\rangle_D,
\end{eqnarray}
where $\sum^N_{m=0}|\gamma_m|^2=1$. It is apparent that $|\varphi_B\rangle$ is a symmetric state, which guarantees that the time-evolved state is also symmetric. The state $|\varphi_B\rangle$ covers several interesting examples in the context of quantum optics and quantum information. For example, the state with $\gamma_m=0$ ($m=0,1,\cdots,N-2$) has been used in Ref.~\cite{Milburn2005} as an initial state of the homogeneous single-qubit CSM with Ising-type interaction to demonstrate the relation between ``entanglement sharing" and single-qubit decoherence. The fully polarized state with $\gamma_m=0$ ($m=0,1,\cdots,N-1$) was used in Ref.~\cite{QB} to study the charging power of a Dicke quantum battery. Another important example is the so-called spin coherent state $|\hat{\Omega}\rangle$ of the spin bath parameterized by the unit vector $\hat{\Omega}=(\sin\theta\cos\phi,\sin\theta\sin\phi,\cos\theta)$. Starting with the fully polarized state $|0\rangle_D$, $|\hat{\Omega}\rangle$ can be obtained by first rotating about the $y$-axis by an angle $\theta$, followed by a rotation about the $z$-axis by an angle $\phi$, i.e., $|\hat{\Omega}\rangle=e^{-iL_z\phi}e^{-iL_y\theta}|0\rangle_D$. It can be shown that $|\hat{\Omega}\rangle$ can be written as~\cite{PRA1972} (see Appendix~\ref{AppA})
\begin{eqnarray}\label{SCSomega}
|\hat{\Omega}\rangle&=&\sum^N_{n=0}Q_n|\frac{N}{2},n-\frac{N}{2}\rangle,
\end{eqnarray}
where
\begin{eqnarray}\label{Qn}
Q_n\equiv\frac{z^n}{(1+|z|^2)^{N/2}}\sqrt{C^n_N},
\end{eqnarray}
with $z\equiv\cot\frac{\theta}{2}e^{-i\phi}$. From Eq.~(\ref{SCSomega}) we identify
\begin{eqnarray}
\gamma_m=Q_{N-m},
\end{eqnarray}
for the spin coherent state $|\hat{\Omega}\rangle$. Note that although $|\hat{\Omega}\rangle$ appears to be a superposition of the Dicke states, the bath spins in this state are actually separable since $|\hat{\Omega}\rangle=\prod^{N}_{j=1}e^{-iT^z_j\phi}e^{-iT^y_j\theta}|+\rangle_j=\prod^{N}_{j=1}\left[\cos\frac{\theta}{2} e^{-i\phi/2}|+\rangle_j+\sin\frac{\theta}{2} e^{i\phi/2}|-\rangle_j\right]$. In this work, we will consider various types of initial states in the form of $|\varphi_B\rangle$.
\subsection{Closed-form time-evolved state for the single-qubit CSM}
\par Before studying the dynamics of the two-qubit CSM, let us first investigate the dynamics of a single qubit coupled uniformly to the spin bath. We will use the Hamiltonian $H^{(1)}_I(t)$ in the interaction picture to derive the equations of motion for the probability amplitudes appearing in a general evolved state of the whole system and solve them analytically.
\par We denote $|\uparrow\rangle$ ($|\downarrow\rangle$) as the up- (down-) state of the qubit and choose the initial state to be a product state,
\begin{eqnarray}
|\psi^{(1)}_0\rangle&=&|\varphi^{(1)}\rangle\otimes|\varphi_B\rangle,
\end{eqnarray}
where
\begin{eqnarray}
|\varphi^{(1)}\rangle&=& \sum_{s=\uparrow,\downarrow }f_{s}|s\rangle,
\end{eqnarray}
is a generic state of qubit $1$ with $ \sum_{s=\uparrow,\downarrow }|f_{s}|^2=1$ and $|\varphi_B\rangle$ is given by Eq.~(\ref{phiB}). To simplify the notations we let $L\equiv N/2$. There are totally $D^{(1)}_{\mathrm{tot}}=2(N+1)$ basis states of the form $|s\rangle|n\rangle_D$ ($s=\uparrow,\downarrow$ and $n=0,\cdots, N$), which are arranged in order as
\begin{eqnarray}
&&|\downarrow\rangle|N\rangle_D,\nonumber\\
&&|\uparrow\rangle|N\rangle_D,|\downarrow\rangle|N-1\rangle_D,\nonumber\\
&&|\uparrow\rangle|N-1\rangle_D,|\downarrow\rangle|N-2\rangle_D,\nonumber\\
&&\vdots\nonumber\\
&&|\uparrow\rangle|1\rangle_D, |\downarrow\rangle|0\rangle_D,\nonumber\\
&&|\uparrow\rangle|0\rangle_D.\nonumber
\end{eqnarray}
\par We write the general time-evolved state in the interaction picture as
\begin{eqnarray}
|\psi^{(1)}_I(t)\rangle=\sum_{s=\uparrow,\downarrow}\sum^N_{n=0}F^{(n)}_s(t)|s\rangle|N-n\rangle_D,
\end{eqnarray}
with initial conditions
\begin{eqnarray}\label{iniF}
F^{(n)}_s(0)=f_s\gamma_{N-n},
\end{eqnarray}
where $\{F^{(n)}_s(t)\}$ are $D^{(1)}_{\mathrm{tot}}$ time-dependent amplitudes determined by the time-dependent Schr\"odinger equation $i\partial_t|\psi^{(1)}_I(t)\rangle=H_I(t)|\psi^{(1)}_I(t)\rangle$. It is easy to see that
\begin{eqnarray}\label{F0N}
F^{(0)}_\downarrow(t)=f_\downarrow \gamma_N,~~F^{(N)}_\uparrow(t)=f_\uparrow \gamma_0
\end{eqnarray}
are constants since $H^{(1)}(t)|\downarrow\rangle|N\rangle_D=H^{(1)}(t)|\uparrow\rangle|0\rangle_D=0$. Due to the conservation of $S^z_1+L_z$, the time evolution takes place separately within each row, yielding the following equations of motion for the two components $F^{(n)}_\downarrow(t)$ and $F^{(n-1)}_\uparrow(t)$ (see Appendix~\ref{Singlem}):
\begin{eqnarray}\label{FF}
i\dot{F}^{(n-1)}_{\uparrow}(t)&=&F^{(n)}_\downarrow(t)g_1x_{n-L-1}e^{- i g'_1 t}e^{ i2g'_1(n-L)t}e^{ i\omega_1t},\nonumber\\
i\dot{F}^{(n)}_{\downarrow}(t)&=&F^{(n-1)}_{\uparrow}(t)g_1 x_{n-L-1}e^{-i\omega_1t} e^{-i2g'_1(n-L)t}e^{ i g'_1 t},\nonumber\\
\end{eqnarray}
where $n=1,2,\cdots,N$ and
\begin{eqnarray}\label{xlm}
x_{m}\equiv\sqrt{(L-m)(L+m+1)}.
\end{eqnarray}
It is shown in Appendix~\ref{ODE} that the analytical solutions of Eqs.~(\ref{F0N}) and (\ref{FF}) with initial conditions given by Eq.~(\ref{iniF}) can be written as
\begin{eqnarray}\label{FFsol}
F^{(n)}_\uparrow(t)&=& e^{\frac{i}{2}b_{n+1}t} ( f_\uparrow \gamma_{N-n} W_{n+1}+f_\downarrow \gamma_{N-n-1} V_{n+1}),\nonumber\\
F^{(n)}_\downarrow(t)&=& e^{-\frac{i}{2}b_nt}(f_\downarrow \gamma_{N-n}W^*_n+f_\uparrow \gamma_{N-n+1} V_n),
\end{eqnarray}
for $n=0,1,\cdots,N$. Here,
\begin{eqnarray}\label{Wn}
W_n \equiv
\cases{
\cos\frac{A_nt}{2}-i\frac{b_n}{A_n}\sin\frac{A_nt}{2}, & $n=1,2,\cdots,N$ \cr
e^{-i\frac{b_{n}t}{2}}, & $n=0$ or $N+1$ \cr
}
\end{eqnarray}
and
\begin{eqnarray}\label{Vn}
V_n \equiv
\cases{
-i\frac{2a_n}{A_n}\sin\frac{A_nt}{2}, & $n=1,2,\cdots,N$ \cr
0, & $n=0$ or $N+1$ \cr
}
\end{eqnarray}
with
\begin{eqnarray}\label{An}
a_n& \equiv& g_1x_{ n-L-1}=a_{N+1-n},\nonumber\\
b_n&\equiv&\omega_1+g'_1[2(n-L)-1],\nonumber\\
A_n&\equiv&\sqrt{b^2_n+4a^2_n}.
\end{eqnarray}
Note that $a_0=a_{N+1}=0$ and $|W_n|^2+|V_n|^2=1$.
\par The Sch\"odinger picture state reads
\begin{eqnarray}\label{psitSch}
|\psi^{(1)}(t)\rangle&=&e^{-i(\omega_1S^z_1+2g'_1S^z_1L_z)t}|\psi^{(1)}_I(t)\rangle\nonumber\\
&=& \sum_{s=\uparrow,\downarrow}\sum^N_{n=0}\tilde{F}^{(n)}_s(t)|s\rangle|N-n\rangle_D,
\end{eqnarray}
where
\begin{eqnarray}\label{FFsoltil}
\tilde{F}^{(n)}_\uparrow(t)&\equiv& e^{-i\frac{1}{2}(b_n+g'_1)t}F^{(n)}_\uparrow(t)\nonumber\\
&=& e^{\frac{i}{2}g'_1t} ( f_\uparrow \gamma_{N-n} W_{n+1}+f_\downarrow \gamma_{N-n-1} V_{n+1}),\nonumber\\
\tilde{F}^{(n)}_\downarrow(t)&\equiv& e^{i\frac{1}{2}(b_n+g'_1)t}F^{(n)}_\downarrow(t)\nonumber\\
&=& e^{ \frac{i}{2}g'_1t}(f_\downarrow \gamma_{N-n}W^*_n+f_\uparrow \gamma_{N-n+1} V_n).
\end{eqnarray}
The analytical form of the time-evolved state $|\psi^{(1)}(t)\rangle$ of the single-qubit CSM offers the basis for analyzing  the reduced dynamics of the single qubit or a spin pair within the spin bath.
\subsubsection{Single-qubit dynamics}
\par The reduced density matrix of the qubit $\vec{S}_1$ can be obtained by tracing out the bath degrees of freedom
\begin{eqnarray}
\rho^{(1)}(t)&=&\mathrm{Tr}_{B}(|\psi^{(1)}(t)\rangle\langle \psi^{(1)}(t)|)\nonumber\\
&=&\sum_{ss'}\left[\sum_n\tilde{F}^{(n)}_s(t)\tilde{F}^{(n)*}_{s'}(t)\right]|s\rangle\langle s'|,
\end{eqnarray}
with the initial density matrix given by
\begin{eqnarray}
\rho^{(1)}(0)&=&\sum_{ss'}f_s f^*_{s'}|s\rangle\langle s'|.
\end{eqnarray}
In the basis $\{|\uparrow\rangle,|\downarrow\rangle\}$, the time-evolved density matrix $\rho^{(1)}(t)$ is connected with $\rho^{(1)}(0)$ via the relations
\begin{eqnarray}\label{rhotrho0}
\rho^{(1)}_{ab}(t)&=&\sum_{cd}Y_{abcd}(t)\rho^{(1)}_{cd}(0),
\end{eqnarray}
where $\rho^{(1)}_{ab}(t)=\langle a|\rho^{(1)}(t)|b\rangle$ ($|a\rangle,|b\rangle=|\uparrow\rangle$ or $|\downarrow\rangle$). Using Eqs.~(\ref{FFsol}) and (\ref{FFsoltil}), we can directly obtain the explicit forms of the coefficients $\{Y_{abcd}(t)\}$, which are listed in Appendix~\ref{AppD}. Equation (\ref{rhotrho0}) provides the complete information for the reduced dynamics of qubit 1.
\par Similarly, we can obtain the reduced density matrix of the spin bath $\rho^{(1)}_B(t)=\sum_{nn'}[\rho^{(1)}_B(t)]_{nn'}|n\rangle_D~_D\langle n'|$ by tracing out the two degrees of freedom of the central spin, where the matrix elements read
\begin{eqnarray}\label{rhoBnn}
[\rho^{(1)}_B(t)]_{nn'}=\sum_{s=\uparrow,\downarrow}\tilde{F}^{(N-n)}_s(t)\tilde{F}^{(N-n')*}_{s}(t),
\end{eqnarray}
with initial condition $[\rho^{(1)}_B(0)]_{nn'}=Q_{N-n}Q_{N-n'}$.
\subsubsection{Reduced dynamics of any two bath spins}\label{single2q}
\par Given the total evolved state $|\psi^{(1)}(t)\rangle$, we are also able to calculate the reduced dynamics of a pair of bath spins. Because of the symmetry of $|\psi^{(1)}(t)\rangle$, we need only to consider the reduced dynamics of any pair of bath spins, say $\vec{T}_1$ and $\vec{T}_2$. We denote $\chi^{(1)}(t)$ as the reduced density matrix of $\vec{T}_1$ and $\vec{T}_2$. The matrix elements of $\chi^{(1)}(t)$ can be calculated in the standard basis $\{|+\rangle_1|+\rangle_2, |+\rangle_1|-\rangle_2,|-\rangle_1|+\rangle_2,|-\rangle_1|-\rangle_2\}$ as~\cite{Bose2009,Wu2010}
\begin{eqnarray}\label{chi12}
\chi^{(1)}_{11}(t) &=&\mathrm{Tr}[\rho^{(1)}_B(t)T^+_1T^-_1T^+_2T^-_2],\nonumber\\
\chi^{(1)}_{22}(t) &=& \mathrm{Tr}[\rho^{(1)}_B(t)T^+_1T^-_1T^-_2T^+_2],\nonumber\\
\chi^{(1)}_{33}(t) &=& \mathrm{Tr}[\rho^{(1)}_B(t)T^-_1 T^+_1T^+_2T^-_2],\nonumber\\
\chi^{(1)}_{44}(t) &=& \mathrm{Tr}[\rho^{(1)}_B(t)T^-_1 T^+_1T^-_2 T^+_2],\nonumber\\
\chi^{(1)}_{12}(t) &=& \mathrm{Tr}[\rho^{(1)}_B(t) T^+_1 T^-_1 T^-_2  ],\nonumber\\
\chi^{(1)}_{13}(t) &=& \mathrm{Tr}[\rho^{(1)}_B(t) T^-_1 T^+_2 T^-_2  ],\nonumber\\
\chi^{(1)}_{14}(t) &=& \mathrm{Tr}[\rho^{(1)}_B(t) T^-_1  T^-_2  ],\nonumber\\
\chi^{(1)}_{23}(t) &=& \mathrm{Tr}[\rho^{(1)}_B(t) T^-_1  T^+_2  ],\nonumber\\
\chi^{(1)}_{24}(t) &=& \mathrm{Tr}[\rho^{(1)}_B(t) T^-_1T^-_2  T^+_2  ],\nonumber\\
\chi^{(1)}_{34}(t) &=& \mathrm{Tr}[\rho^{(1)}_B(t) T^-_1T^+_1  T^-_2  ].
\end{eqnarray}
Other off-diagonal matrix elements can be obtained from the Hermitian property of the density matrix. To obtain the matrix elements $\chi^{(1)}_{ij}(t)$, we have to obtain the matrix representations of the operators $T^+_1T^-_1T^+_2T^-_2$, etc., in the Dicke basis. A straightforward analysis based on Eq.~(\ref{Dickestate}) gives the following nonvanishing matrix elements (see Appendix~\ref{AppE})
\begin{eqnarray}\label{mTTm}
~_D\langle m|T^+_1T^-_1T^+_2T^-_2|m\rangle_D&=& C^m_{N-2}/C^m_N,\nonumber\\
~_D\langle m|T^+_1T^-_1T^-_2T^+_2|m\rangle_D&=& ~_D\langle m|T^-_1T^+_1T^+_2T^-_2|m\rangle_D\nonumber\\
&=&~_D\langle m|T^-_1T^+_2|m\rangle_D\nonumber\\
&=&C^{m-1}_{N-2}/C^m_N,\nonumber\\
~_D\langle m|T^-_1T^+_1T^-_2T^+_2|m\rangle_D&=&C^{m-2}_{N-2}/C^m_N,\nonumber\\
~_D\langle m+1|T^+_1T^-_1T^-_2|m\rangle_D&=& ~_D\langle m+1|T^-_1T^+_2T^-_2|m\rangle_D\nonumber\\
&=& C^m_{N-2}/\sqrt{C^{m+1}_N C^m_N},\nonumber\\
~_D\langle m+1|T^-_1T^-_2T^+_2|m\rangle_D&=&~_D\langle m+1|T^-_1 T^+_1T^-_2|m\rangle_D\nonumber\\
&=& C^{m-1}_{N-2}/\sqrt{C^{m+1}_N C^m_N},\nonumber\\
~_D\langle m+2|T^-_1T^-_2|m\rangle_D&=&C^{m}_{N-2}/\sqrt{C^{m+2}_N C^m_N},
\end{eqnarray}
where we used the convention $C^i_j=0$ if $i<0$ or $i>j$.
\par By combining Eq.~(\ref{chi12}) with Eq.~(\ref{mTTm}), we finally obtain the reduced density matrix of $\vec{T}_1$ and $\vec{T}_2$,
\begin{eqnarray}
\chi^{(1)} (t) &=&  \left(
                       \begin{array}{cccc}
                         p^{(1)}_1 & p^{(1)}_3 & p^{(1)}_3 & p^{(1)}_4 \\
                         p^{(1)*}_3 & p^{(1)}_2 & p^{(1)}_2 & p^{(1)}_5 \\
                         p^{(1)*}_3 & p^{(1)}_2 & p^{(1)}_2 & p^{(1)}_5 \\
                         p^{(1)*}_4 & p^{(1)*}_5 & p^{(1)*}_5 & 1-p^{(1)}_1-2p^{(1)}_2 \\
                       \end{array}
                     \right),\nonumber\\
\end{eqnarray}
where
\begin{eqnarray}\label{pppp}
p^{(1)}_1(t)&=&\sum^{N-2}_{m=0}[\rho^{(1)}_B(t)]_{mm }C^m_{N-2}C^m_N,\nonumber\\
p^{(1)}_2(t)&=& \sum^{N-1}_{m=1}[\rho^{(1)}_B(t)]_{mm }C^{m-1}_{N-2}/C^m_N,\nonumber\\
p^{(1)}_3(t)&=&\sum^{N-2}_{m=0}[\rho^{(1)}_B(t)]_{m,m+1}C^m_{N-2}/ \sqrt{C^{m+1}_NC^m_N},\nonumber\\
p^{(1)}_4(t)&=&\sum^{N-2}_{m=0}[\rho^{(1)}_B(t)]_{m,m+2} C^m_{N-2}/\sqrt{C^{m+2}_NC^m_N},\nonumber\\
p^{(1)}_5(t)&=& \sum^{N-1}_{m=1} [\rho^{(1)}_B(t)]_{m,m+1} C^{m-1}_{N-2}/\sqrt{C^{m+1}_NC^m_N}.
\end{eqnarray}
\subsection{Time-evolved state for the two-qubit CSM}\label{twoqwf}
\par We now consider the dynamics governed by Hamiltonian (\ref{H}), which describes two coupled qubits interacting with a common spin bath via uniform system-bath coupling. We still assume a separable initial state
\begin{eqnarray}
|\psi_0\rangle&=&|\varphi\rangle|\varphi_B\rangle,
\end{eqnarray}
where
\begin{eqnarray}
|\varphi\rangle&=& \sum_{s_1,s_2=\uparrow,\downarrow }A_{s_1,s_2}|s_1,s_2\rangle,
\end{eqnarray}
with $\sum_{s_1,s_2=\uparrow,\downarrow }|A_{s_1,s_2}|^2=1$. There are totally $D_{\mathrm{tot}}=4(N+1)$ basis states of the form $|s_1,s_2\rangle|n\rangle_D$, and we arrange them as
\begin{eqnarray}\label{basis2q}
&&|\downarrow\downarrow\rangle|N\rangle_D, \nonumber\\
&&|\uparrow\downarrow\rangle|N\rangle_D, |\downarrow\uparrow\rangle|N\rangle_D, |\downarrow\downarrow\rangle|N-1\rangle_D\nonumber\\
&&|\uparrow\uparrow\rangle|N\rangle_D, |\uparrow\downarrow\rangle|N-1\rangle_D, |\downarrow\uparrow\rangle|N-1\rangle_D, |\downarrow\downarrow\rangle|N-2\rangle_D\nonumber\\
&&\vdots\nonumber\\
&&|\uparrow\uparrow\rangle|2\rangle_D, |\uparrow\downarrow\rangle|1\rangle_D, |\downarrow\uparrow\rangle|1\rangle_D, |\downarrow\downarrow\rangle|0\rangle_D\nonumber\\
&&|\uparrow\uparrow\rangle|1\rangle_D, |\uparrow\downarrow\rangle|0\rangle_D, |\downarrow\uparrow\rangle|0\rangle_D,\nonumber\\
&&|\uparrow\uparrow\rangle|0\rangle_D.
\end{eqnarray}
There are $2L+3=N+3$ rows in the above equation and it is easy to see that $H_I(t)|\downarrow\downarrow\rangle|N\rangle_D=H_I(t)|\uparrow\uparrow\rangle|0\rangle_D=0$. The time-evolved state in the interaction picture can be expanded in terms these basis states as
\begin{eqnarray}
|\psi_I(t)\rangle&=&\sum_{s_1,s_2=\uparrow,\downarrow}\sum^N_{n=0} G^{(n)}_{s_1s_2}(t)|s_1,s_2\rangle|N-n\rangle_D,
\end{eqnarray}
with initial conditions $G^{(n)}_{s_1s_2}(0)=A_{s_1s_2}\gamma_{N-n}$. Due to the conservation of the magnetization $M=S^z_1+S^z_2+L_z$, the Hamiltonian $H$ is block diagonal in the basis given by Eq.~(\ref{basis2q}), and we can separately treat the time evolution within subspaces spanned by states in each row. As before, application of the Schr\"odinger equation to the state $|\psi_I(t)\rangle$ leads to the following sets of equations of motion
\begin{eqnarray}\label{c++}
&&i\dot{G}^{(n-1)}_{\uparrow\uparrow}= G^{(n)}_{\uparrow\downarrow}g_2x_{n-L-1}e^{ i\omega_2t}e^{ -ig'_{12} t}e^{ iJ' t}e^{ i 2g'_2 (n-L-1)t}\nonumber\\
&&+G^{(n)}_{\downarrow\uparrow}g_1x_{n-L-1}e^{ig'_{12} t}e^{ i\omega_1t}e^{i 2g'_1 (n-L-1)t}e^{iJ' t},\nonumber\\
&&i\dot{G}^{(n)}_{\uparrow\downarrow}=G^{(n)}_{ \downarrow\uparrow}J  e^{i 2g'_{12}(n-L)t}e^{i\omega_{12}t}\nonumber\\
&&+G^{(n-1)}_{\uparrow\uparrow}g_2 x_{n-L-1}e^{ ig'_{12} t}e^{-i\omega_2t}e^{-i 2g'_2 (n-L-1)t}e^{-iJ't}\nonumber\\
&&+G^{(n+1)}_{\downarrow\downarrow}g_1x_{n-L}e^{ -ig'_{12} t}e^{ i\omega_1t}e^{i 2g'_1 (n-L+1)t}e^{-iJ' t},\nonumber\\
&&i\dot{G}^{(n)}_{\downarrow\uparrow}=G^{(n)}_{\uparrow\downarrow}Je^{-i2g'_{12}(n-L)t}e^{-i\omega_{12}t}\nonumber\\
&&+G^{(n-1)}_{ \uparrow\uparrow}g_1 x_{n-L-1}e^{- ig'_{12} t}e^{-i\omega_1t}e^{-i 2g'_1 (n-L-1)t}e^{-iJ' t}\nonumber\\
&&+G^{(n+1)}_{\downarrow\downarrow}g_2x_{n-L}e^{ ig'_{12} t}e^{ i\omega_2t}e^{ i 2g'_2 (n-L+1)t}e^{ -iJ' t},\nonumber\\
&&i\dot{G}^{(n+1)}_{\downarrow\downarrow}= G^{(n)}_{\uparrow\downarrow}g_1 x_{n-L}e^{-i\omega_1t}e^{ ig'_{12} t}e^{iJ' t}e^{-i 2g'_1 (n-L+1)t}\nonumber\\
&&+G^{(n)}_{\downarrow\uparrow}g_2 x_{n-L}e^{- ig'_{12} t}e^{-i\omega_2t}e^{-i 2g'_2(n-L+1)t}e^{iJ' t},
\end{eqnarray}
for $n=1,2,\cdots,N-1$. The above set of equations can also incorporate the cases of $n=0$ and $n=N$, with the understanding that $G^{(-1)}_{\uparrow\uparrow}=G^{(N+1)}_{\downarrow\downarrow}\equiv0$. In contrast to the single-qubit dynamics, these coupled equations of motion can no longer be solved analytically.
\par The Schr\"odinger picture state is
\begin{eqnarray}
|\psi(t)\rangle&=&e^{-iH_0t}|\psi_I(t)\rangle\nonumber\\
&\equiv&\sum_{s_1,s_2=\uparrow,\downarrow}\sum^N_{n=0} \tilde{G}^{(n)}_{s_1s_2}(t)|s_1,s_2\rangle|N-n\rangle_D,
\end{eqnarray}
where
\begin{eqnarray}
 \tilde{G}^{(n)}_{s_1s_2}(t)&=& G^{(n)}_{s_1s_2}(t)e^{-\frac{i}{2}(v_1\omega_1+v_2\omega_2+J'v_1v_2)t}\nonumber\\
 &&e^{-i(g'_1v_1+g'_2v_2)(n-L)t},
\end{eqnarray}
with $v_{i}=1$ or $-1$ for $s_{i}=\uparrow$ or $\downarrow$.
\subsubsection{Two-qubit dynamics}
\par The reduced density matrix (in the standard basis) of the two central qubits can be obtained by tracing out the bath degrees of freedom
\begin{eqnarray}
\rho(t)&=&\mathrm{Tr}_B[|\psi(t)\rangle\langle\psi(t)|]\nonumber\\
&=&  \sum_{s_1,s_2 } \sum_{s'_1,s'_2 } \sum^{N}_{n=0} \tilde{G}^{(n)}_{s_1s_2}(t)\tilde{G}^{(n)*}_{s'_1s'_2}(t)|s_1s_2\rangle \langle s'_1s'_2|\nonumber\\
&=& \left(
             \begin{array}{cccc}
               C_{\uparrow\uparrow;\uparrow\uparrow}  &  C_{\uparrow\uparrow;\uparrow\downarrow} &  C_{\uparrow\uparrow;\downarrow\uparrow} &  C_{\uparrow\uparrow;\downarrow\downarrow} \\
               C_{\uparrow\downarrow;\uparrow\uparrow}  &  C_{\uparrow\downarrow;\uparrow\downarrow} &  C_{\uparrow\downarrow;\downarrow\uparrow} &  C_{\uparrow\downarrow;\downarrow\downarrow} \\
               C_{\downarrow\uparrow;\uparrow\uparrow}  &  C_{\downarrow\uparrow;\uparrow\downarrow} &  C_{\downarrow\uparrow;\downarrow\uparrow} &  C_{\downarrow\uparrow;\downarrow\downarrow} \\
               C_{\downarrow\downarrow;\uparrow\uparrow}  &  C_{\downarrow\downarrow;\uparrow\downarrow} &  C_{\downarrow\downarrow;\downarrow\uparrow} &  C_{\downarrow\downarrow;\downarrow\downarrow} \\
             \end{array}
           \right),
\end{eqnarray}
where
\begin{eqnarray}
&&C_{s_1s_2;s'_1s'_2}(t)=\sum^{N}_{n=0} \tilde{G}^{(n)}_{s_1s_2}(t)\tilde{G}^{(n)*}_{s'_1s'_2}(t).
\end{eqnarray}
\par We can similarly obtain the reduced density matrix of the spin bath $\rho_B(t)=\sum_{nn'}[\rho_B(t)]_{nn'}|n\rangle_D~_D\langle n'|$ by tracing out the four degrees of freedom of the central two-qubit system, where
\begin{eqnarray}\label{rhoBnn2q}
[\rho^{(1)}_B(t)]_{nn'}=\sum_{s_1s_2}\tilde{G}^{(N-n)}_{s_1s_2}(t)\tilde{G}^{(N-n')*}_{s_1s_2}(t),
\end{eqnarray}
with initial condition $[\rho_B(0)]_{nn'}=Q_{N-n}Q_{N-n'}$.
\subsubsection{Reduced dynamics of two bath spins}
\par Following almost the same analysis as in Sec.\ref{single2q}, we can get the reduced density matrix $\chi(t)$ of $\vec{T}_1$ and $\vec{T}_2$:
\begin{eqnarray}
\chi (t) &=&  \left(
                       \begin{array}{cccc}
                         p_1 & p_3 & p_3 & p_4 \\
                         p^{*}_3 & p_2 & p_2 & p_5 \\
                         p^{ *}_3 & p_2 & p_2 & p_5 \\
                         p^{ *}_4 & p^{ *}_5 & p^{ *}_5 & 1-p_1-2p_2 \\
                       \end{array}
                     \right),
\end{eqnarray}
where the $p$'s are obtained by replacing the $\rho^{(1)}_B(t)$ with $\rho_B(t)$ in Eq.~(\ref{pppp}).
\section{Reduced dynamics of the central system}\label{SecIII}
\par Using the formalism developed in the last section, we now study the reduced dynamics of the central systems with the spin bath prepared in the spin coherent state. For the single-qubit CSM, we will provide an interpretation to the collapse-revival phenomena observed in previous literatures. We then turn to study the entanglement and coherence dynamics of the two central qubits in the two-qubit CSM, with the qubits coupled to individual or common spin baths.
\subsection{Single-qubit CSM}
\par We suppose that the central spin is initially in its up-state, i.e., $f_\uparrow=1$ and $f_{\downarrow}=0$, then the polarization $\langle S^z_1(t)\rangle$ can be expressed as
\begin{eqnarray}
\langle S^z_1(t)\rangle&=&\frac{1}{2}\left[\rho^{(1)}_{\uparrow\uparrow}(t)-\rho^{(1)}_{\downarrow\downarrow}(t)\right]\nonumber\\
&=&\frac{1}{2}\sum^{N}_{n=0}|Q_{n}|^2\frac{b^2_{n+1}+4a^2_{n+1}\cos A_{n+1}t}{A^2_{n+1}},
\end{eqnarray}
which is consistent with the result in Ref.~\cite{Guan2019}.
\par The off-diagonal element of $\rho^{(1)}(t)$ characterizes the coherence of the qubit and is given by $\langle S^+_1(t)\rangle=\rho^{(1)}_{\downarrow\uparrow}(t)$, yielding
\begin{eqnarray}\label{Sx}
\langle S^x_1(t)\rangle&=&\sum^N_{n=1}\frac{ Q_n Q_{n-1} a_nb_{n+1}}{A_nA_{n+1}}\nonumber\\
&&\left[\cos\frac{(A_n-A_{n+1})t}{2}-\cos\frac{(A_n+A_{n+1})t}{2}\right],\nonumber\\
\end{eqnarray}
and
\begin{eqnarray}\label{Sy}
\langle S^y_1(t)\rangle&=&-\sum^N_{n=1}\frac{ Q_n Q_{n-1} a_n}{A_n}\nonumber\\
&&\left[\sin\frac{(A_n+A_{n+1})t}{2}+\sin\frac{(A_n-A_{n+1})t}{2}\right].\nonumber\\
\end{eqnarray}
We also monitor the purity dynamics defined by
\begin{eqnarray}\label{Pt}
P(t)=\frac{1}{2}+2\sum_{i=x,y,z}\langle S^i_1(t)\rangle^2.
\end{eqnarray}
Below we will focus on two limiting cases, i.e., the fully anisotropic ($XX$-type coupling) and fully isotropic ($XXX$-type coupling) limit.
\subsubsection{$XX$-type coupling: $g'_1=0$}
\par First, we would like to mention that it is shown recently that the homogeneous $XX$ central spin model is exactly solvable via the Bethe ansatz for $s_1=1/2$ and arbitrary $\{t_i\}$~\cite{PRB2020} ($t_i$ is the quantum number of the $i$th bath spin $\vec{T}_i$). The situation studied here amounts to the special case of $t_i=1/2,~\forall i$. Note that collapse and revival phenomena in such a model has been numerically revealed in Refs.~\cite{CRcat,PRA2014}. Here, we will provide a quantitative explanation of the observed collapse-revival phenomena.
\par
\begin{figure}
\includegraphics[width=.53\textwidth]{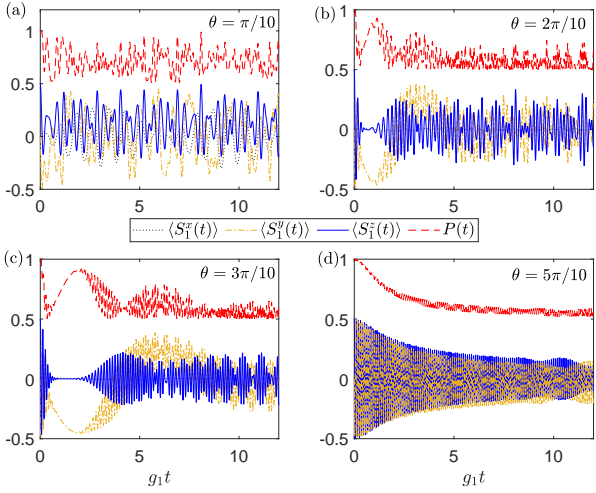}
\caption{Time evolution of the three components $\langle S^\alpha_1(t)\rangle$ and the purity $P(t)$ for a single central spin ($\vec{S}_1$) initially prepared in its up state. Results for $\theta=\pi/10,~2\pi/10,~3\pi/10$, and $5\pi/10$ are presented to show the dependence of the dynamics on the bath initial conditions. Parameters: $N=60,~\omega_1/g_1=1,~g'_1/g_1=0$.}
\label{Fig2}
\end{figure}

 We plot in Fig.~\ref{Fig2} the dynamics of the three components $\langle S^\alpha_1(t)\rangle$ and the purity $P(t)$ starting from the upper state $|\uparrow\rangle$ of qubit 1 for four different values of $\theta$ ranging from $\theta=0$ to $\theta=\pi/2$. It can be seen that for $\theta=\pi/10$ all quantities oscillate randomly and no features are developed as time evolves [Fig.~\ref{Fig2}(a)]. In the other extreme case with $\theta=\pi/2$, the purity $P(t)$ decreases gradually with slight oscillations and approaches a completely mixed state with $P=1/2$ in the long-time limit [Fig.~\ref{Fig2}(d)]. In contrast, for $3\pi/10$, we first observe an abrupt collapse of the qubit state at a short-time scale, which is followed, to a good approximation, by a revival of some pure state with a high purity at $g_1t\approx 2.1$. The approximate recreation of the state vector is accompanied by a collapse in $\langle S^z_1(t)\rangle$ and by nearly vanishing $\langle S^x_1(t)\rangle$, indicating that the reached qubit state is approximately the state pointing along the $-y$ direction.
\par
\begin{figure}
\includegraphics[width=.49\textwidth]{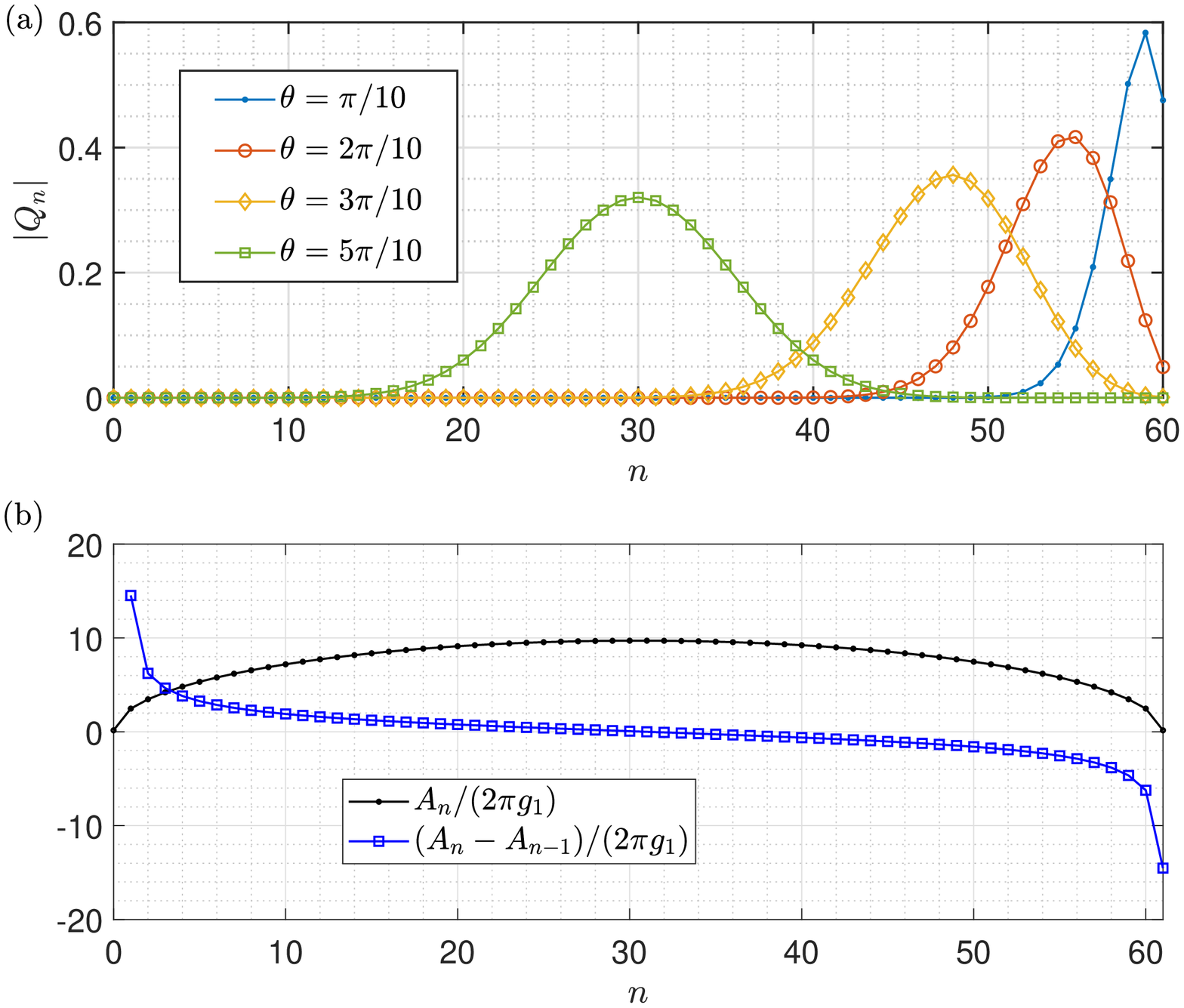}
\includegraphics[width=.49\textwidth]{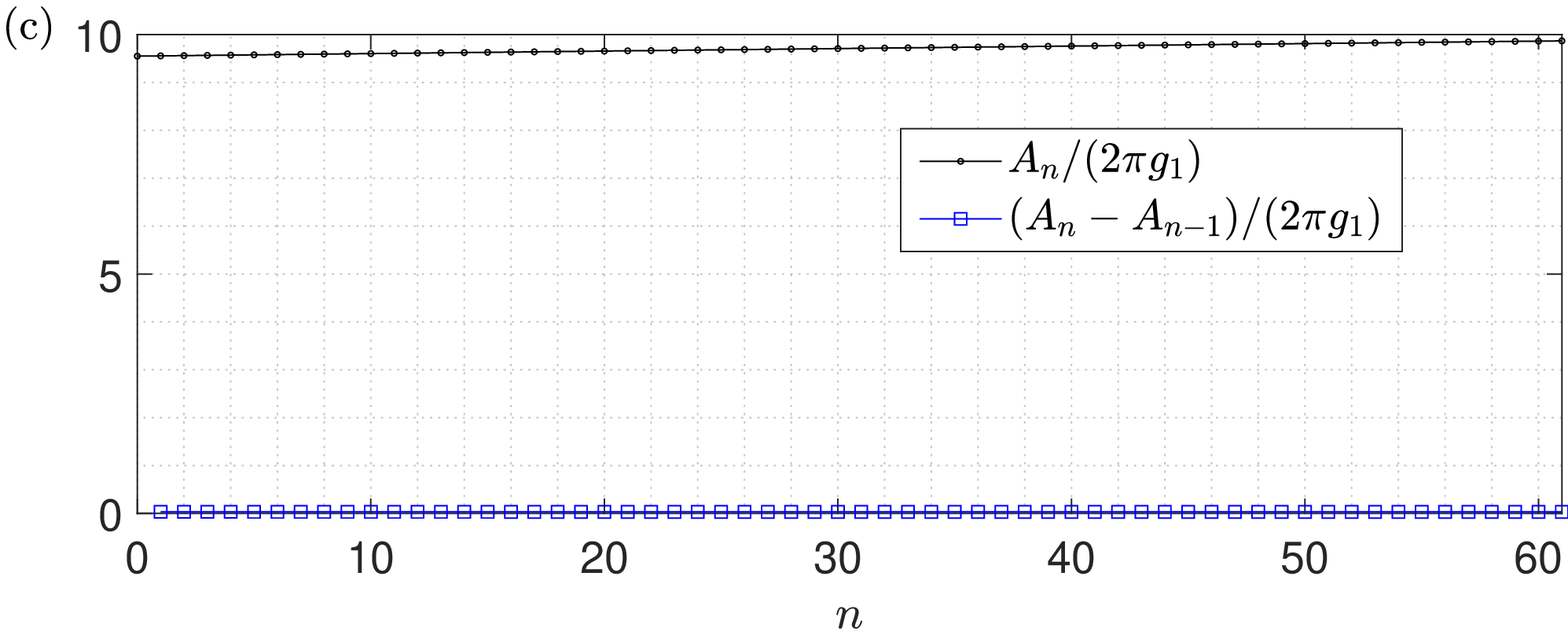}
\caption{(a) The dependence of the coefficients $Q_n(\theta,\phi)$ [Eq.~(\ref{Qn})] on $\theta$. Note that $|Q_n(\theta,\phi)|^2=|Q_{N-n}(\pi-\theta,\phi)|^2$. (b) and (c) show the Rabi frequency $A_n$ [Eq.~(\ref{An})] and the neighboring frequency difference $A_n-A_{n-1}$ for $g'_1=0$ and $g'_1=g_1$, respectively. Parameters: $N=60$, $\phi=0$, and $\omega_1/g_1=1$.}
\label{Fig3}
\end{figure}

 The revival of the qubit state and collapse of the $\langle S^z_1(t)\rangle$ profile found in the present model with $g'_1=0$ are very similar to the phenomenon observed in the Jaynes-Cummings model~\cite{PRL1990} and the relationship between the two models was discussed in Refs.~\cite{CRcat,Guan2019}. To understand the above observations in the context of a spin bath, we note from Eqs.~(\ref{FFsol}), (\ref{Wn}), (\ref{Vn}), and (\ref{psitSch}) that the time-evolved state $|\psi^{(1)}(t)\rangle$ depends on both the coefficients $\{Q_n\}$ characterizing the bath initial condition and the Rabi frequencies $\{A_n\}$. In Fig.~\ref{Fig3}(a) we plot $|Q_n|$ for several values of $\theta$ and find that for each one there is some $n=n_{\max}$ around which $|Q_n|$ is significantly different from zero. Suppose at some time $t_r$ a revival of the qubit pure state is approximately achieved, then the state $|\psi^{(1)}(t)\rangle$ will have the form
\begin{eqnarray}
|\psi^{(1)}(t_r)\rangle\approx\sum_{n\approx n_{\max}}[\sum_{s=\uparrow,\downarrow}\tilde{F}^{(n)}_s(t_r)|s\rangle]|n-L\rangle,
\end{eqnarray}
where the ratio between $\tilde{F}^{(n)}_\uparrow(t_r)$ and $\tilde{F}^{(n)}_\downarrow(t_r)$,
\begin{eqnarray}\label{FFratio}
\frac{\tilde{F}^{(n)}_\uparrow(t_r)}{\tilde{F}^{(n)}_\downarrow(t_r)}&=& \frac{ Q_{n} W_{n+1}(t_r) }{  Q_{n-1} V_n(t_r) }\nonumber\\
&=&\frac{Q_n}{Q_{n-1}}\frac{\cos\frac{A_{n+1}t_r}{2}-i\frac{b_{n+1}}{A_{n+1}}\sin\frac{A_{n+1}t_r}{2}}{-i\frac{2a_n}{A_n}\sin\frac{A_nt_r}{2}},
\end{eqnarray}
should be almost independent of $n$ for $n\approx n_{\max}$.
\par Now observing that for $g'_1=0$ we have $a_n=g_1\sqrt{(2L-n+1)n}$ and $b_n=\omega_1$. Suppose $g_1\sim \omega_1$ and note that $n_{\max}\gg 1$, then $a_n\gg b_n$, so that $A_n\approx 2a_n$. The second term in the numerator of Eq.~(\ref{FFratio}) can thus be neglected, giving
\begin{eqnarray}
\frac{\tilde{F}^{(n)}_\uparrow(t_r)}{\tilde{F}^{(n)}_\downarrow(t_r)}\approx\frac{Q_n}{Q_{n-1}}\frac{\cos\frac{A_{n+1}t_r}{2} }{- i \sin\frac{A_nt_r}{2}}\approx i\frac{\cos\frac{A_{n+1}t_r}{2} }{  \sin\frac{A_nt_r}{2}},
\end{eqnarray}
where we used the fact that $Q_n\approx Q_{n-1}$ for $n$ around $n_{\max}$. From Fig.~\ref{Fig3}(b) we see that $A_n$ decreases with increasing $n$ for $n\geq L$. At the time satisfying
\begin{eqnarray}
t_r= \frac{\pi}{A_n -A_{n+1}},
\end{eqnarray}
we have $\cos\frac{A_{n+1}t_r}{2} = \sin\frac{A_nt_r}{2}$, and hence $\tilde{F}^{(n)}_\uparrow(t_r)/\tilde{F}^{(n)}_\downarrow(t_r)\approx i$, which gives an approximately separable state
\begin{eqnarray}\label{psisep}
|\psi^{(1)}(t_r)\rangle\approx\frac{1}{\sqrt{2}}(|\uparrow\rangle-i|\downarrow\rangle)\sum_{n\approx n_{\max}}\tilde{F}^{(n)}_\downarrow(t_r)|n-L\rangle.
\end{eqnarray}
The qubit part of $|\psi^{(1)}(t_r)\rangle$ is just $|-\hat{y}\rangle$, i.e., the state pointing along the $-y$ direction. For the typical case with $\theta=3\pi/10$, we have $n_{\max}=48$ and $(A_{n_{\max}}-A_{n_{\max}+1})\approx 1.47$, so that $g_1 t_r\approx 2.14$, in consistent with the numerical results presented in Fig.~\ref{Fig2}(c). For $\theta=\pi/10$ and $5\pi/10$, the frequency difference $A_n-A_{n+1}$ for $n_{\max}=59$ ($n_{\max}=30$) is so large (small) [see Fig.~\ref{Fig3}(b)] that the revival of $P(t)$ occurs at very short (long) time scales [Figs.~\ref{Fig2}(a),(d)].
\begin{figure}
\includegraphics[width=.53\textwidth]{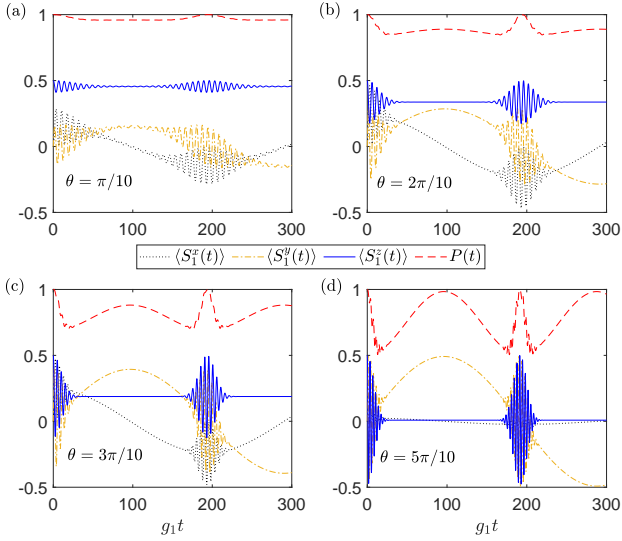}
\caption{The same as in Fig.~\ref{Fig2}, but for $g'_1/g_1=1$.}
\label{Fig4}
\end{figure}
\subsubsection{$XXX$-type coupling: $g'_1=g_1$}
\par In Fig.~\ref{Fig4}, we plot the dynamics of $\langle S^\alpha_1(t)\rangle$ and $P(t)$ for qubit 1 coupled to the spin bath via $XXX$-type coupling. We assume $g_1\approx\omega_1$, then $b_n\approx 2g_1  (n-L) $, giving
\begin{eqnarray}
A_n&\approx & 2g_1\sqrt{  L^2 + n}\approx 2g_1 L\left(1+\frac{n}{2L^2}\right).
\end{eqnarray}
So $A_n$ increases linearly with increasing $n$ with a small slope $g_1/L$ for large $L$ [Fig.~\ref{Fig3}(c)]. By noting that the frequency difference $A_{n+1}-A_n\approx g_1/L$ is approximately independent of $n$, revivals of the purity are expected to occur at times satisfying $t_r= (2m+1)L\pi/g_1$ [Fig.~\ref{Fig4}(d)].
\subsection{Two-qubit CSM}\label{SecIV}
\par We now turn to study the reduced dynamics of two qubits homogeneously coupled to spin baths. We will consider two situations, i.e., the two qubits couple to individual baths and to a common bath.
\subsubsection{Disentanglement of two noninteracting qubits coupled to individual spin baths}
\par The composite system consisting of two qubits interacting with their own environments provides an interesting setup to study the dynamics of entanglement~\cite{erbly,nonmark,Tanimura2010}. The single-qubit reduced dynamics obtained in the previous section can be directly used to obtain the reduced dynamics of two such copies, provided that there is no direct interaction between the two~\cite{nonmark}. The Hamiltonian of such a composite system reads [see Fig.~\ref{Fig1}(b)]
\begin{eqnarray}\label{H2q}
H_{\mathrm{ind}}&=&\sum_{\alpha=1,2}H^{(\alpha)}_S+H^{(\alpha)}_{SB},\nonumber\\
H^{(\alpha)}_S&=&\omega_\alpha S^z_\alpha,\nonumber\\
H^{(\alpha)}_{SB}&=&2\sum^N_{j=1}\left[g_\alpha(S^x_\alpha T^{(\alpha)x}_j+S^y_\alpha T^{(\alpha)y}_j)+g'_\alpha S^z_\alpha T^{(\alpha)z}_{j}\right].\nonumber\\
\end{eqnarray}
The two baths are independent in the sense that any two operators belonging to distinct copies commute with each other. The entanglement dynamics of a pair of two-level atoms coupled to individual photonic baths is studied in Refs.~\cite{erbly,nonmark}.
\par Let $\rho(t)$ be the reduced density matrix of the two uncoupled qubits. We assume a separable initial state for the whole system
\begin{eqnarray}\label{rho2qtot}
\rho_{\mathrm{tot}}(0)=\rho(0)\otimes\rho^{(1)}_B\otimes\rho^{(2)}_B,
\end{eqnarray}
where $\rho^{(\alpha)}_B$ is the bath initial state of bath $\alpha$. For the sake of consistency, we still choose the spin coherent state as the bath initial state, i.e., $\rho^{(\alpha)}_B=|\hat{\Omega}^{(\alpha)}\rangle\langle \hat{\Omega}^{(\alpha)}|$. The reduced density matrix $\rho(t)$ can be expressed in the standard basis $\{|\uparrow\uparrow\rangle,|\uparrow\downarrow\rangle,|\downarrow\uparrow\rangle,|\downarrow\downarrow\rangle\}$ as~~\cite{nonmark}
\begin{eqnarray}\label{rho2q}
\rho(t)_{aa',bb'}=\sum_{cc'dd'}Y^{(1)}_{abcd}(t)Y^{(2)}_{a'b'c'd'}(t)\rho_{cc',dd'}(0),
\end{eqnarray}
where $Y^{(1)}_{abcd}(t)$ is determined by the single-qubit dynamics given by Eq.~(\ref{rhotrho0}).
\par We focus on the disentanglement dynamics of the maximally entangled state $|\varphi\rangle=(|\uparrow\downarrow\rangle+ |\downarrow\uparrow\rangle)/\sqrt{2}$ (so that $\rho(0)=|\varphi\rangle\langle\varphi|$). We use Wootters's concurrence~\cite{Wootters} to measure the entanglement between the two qubits, which is defined as
\begin{eqnarray}
\mathcal{C}(t)=\max\{0,\sqrt{\lambda_1}-\sqrt{\lambda_2}-\sqrt{\lambda_3}-\sqrt{\lambda_4}\},
\end{eqnarray}
where $\lambda_i$ are the positive eigenvalues of the matrix $\rho(\sigma_y\otimes\sigma_y)\rho^*(\sigma_y\otimes\sigma_y)$ arranged in decreasing order.
\begin{figure}
\includegraphics[width=.54\textwidth]{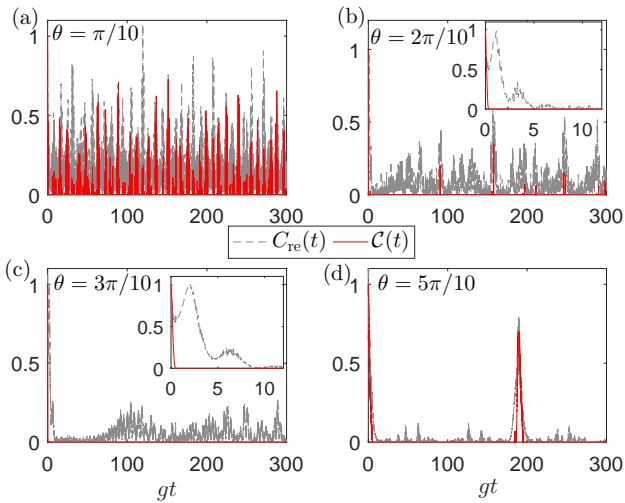}
\caption{Time evolution of the relative entropy of coherence $C_{\rm{re}}(t)$ (dashed gray curves) and the concurrence $\mathcal{C}(t)$ (red curves) for two qubits independently coupled to their own spin bath via the $XX$-type coupling [see Eq.~(\ref{H2q})]. The initial state of the two qubits is chosen as the maximally entangled state $|\varphi\rangle=\frac{1}{\sqrt{2}}(|\uparrow\downarrow\rangle+|\downarrow\uparrow\rangle)$ and we choose $\theta_1=\theta_2=\theta$, $g_1=g_2=g$, $\omega_1=\omega_2=\omega$, and $g'_1=g'_2=g'$ for the two baths. The insets in (b) and (c) shows the corresponding short-time dynamics up to $gt=12$. Parameters: $N=60$, $\omega/g=1$, and $g'/g=0$.}
\label{Fig5}
\end{figure}
\par We are also interested in the time evolution of the coherence of the two-qubit state, which is attributed to the off-diagonal elements of the two-qubit reduce density matrix with respect to a selected reference basis. Here, we will use the relative entropy of coherence~\cite{Plenio} as a measure of coherence
\begin{eqnarray}
C_{\rm{re}}(t)=S(\rho_{\rm{diag}}(t))-S(\rho(t)),
\end{eqnarray}
where $S(\rho)=-\mathrm{Tr}(\rho\ln \rho)$ is the von Neumann entropy and $\rho_{\rm{diag}}$ is the density matrix from $\rho$ by removing all off-diagonal elements.
\par In Fig.~\ref{Fig5}, we present the time evolution of $C_{\rm{re}}(t)$ and $\mathcal{C}(t)$ for the $XX$-type coupling and various values of $\theta$. Similar to the single-qubit polarization dynamics, both $C_{\rm{re}}(t)$ and $\mathcal{C}(t)$  oscillate irregularly for $\theta=\pi/10$. However, for $\theta=2\pi/10$ we observe an abrupt short decay of the concurrence to zero, which is followed by several revivals peaked at nearly discrete time slices. Similar behaviors are observed for $\theta=3\pi/10$ and $5\pi/10$. The relative entropy is still present even for vanishing concurrence, indicating that the state of the two qubits might be coherent even in the absence of entanglement between the two. Note that the revival of the concurrence is accompanied by sudden increase in the relative entropy, so the coherence and entanglement are somehow related. Interestingly, we find that the coherence behaves similarly to the single-qubit purity shown in Fig.~\ref{Fig3} [by comparing, e.g., Fig.~\ref{Fig3}(b), (c) with the insets in Fig.~\ref{Fig5}(b), (c)], indicating the two quantities are positively correlated.

\par
\begin{figure}
\includegraphics[width=.54\textwidth]{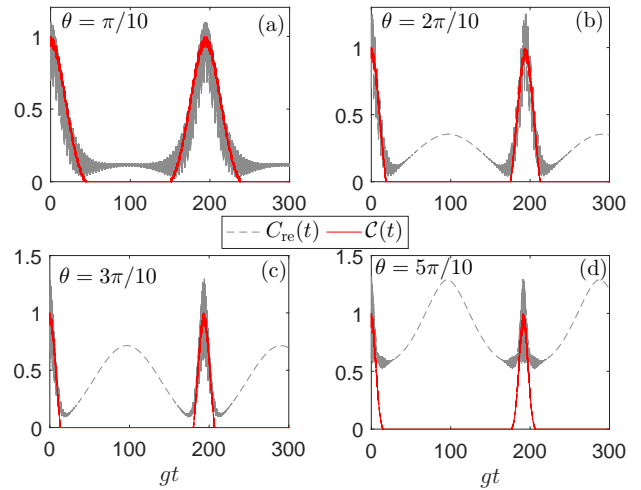}
\caption{The same as in Fig.~\ref{Fig5}, but for the $XXX$-type coupling with $g'/g=1$.}
\label{Fig6}
\end{figure}

 Figure~\ref{Fig6} shows the dynamics of $C_{\rm{re}}(t)$ and $\mathcal{C}(t)$ for the $XXX$-type coupling. We see that the two-qubit entanglement decays to zero in a short time for all values of $\theta$ considered. After a period of complete disentanglement, a revival of nearly maximal entanglement occurs at $gt\approx N\pi$ (we assume $g_1=g_2=g$), a time at which the single-qubit Rabi oscillation also revives (see Fig.~\ref{Fig4}). Note that similar phenomenon of entanglement disappearance and revival is observed in the non-Markovian dynamics of two noninteracting qubits independently coupled to zero-temperature bosonic reservoirs~\cite{nonmark}. Actually, the interaction of quantum spins with a spin bath can in general lead to non-Markovian behavior in the dynamics~\cite{Lidar2007,Breuer2008,Lorenzo2013}. We also observe a positive correlation between the coherence and the single-qubit purity shown in Fig.~\ref{Fig4}.
\subsubsection{Entanglement dynamics of two qubits coupled to a common spin bath}
\par We now consider the reduced two-qubit dynamics governed by Hamiltonian (\ref{H}), which describes two coupled qubits interacting with a common spin bath via uniform system-bath coupling. The following numerical analysis are based on the results obtained in Sec.~\ref{twoqwf}.
\par We assume the bath is prepared in the spin coherent state $|\hat{\Omega}\rangle$
\begin{figure}
\includegraphics[width=.54\textwidth]{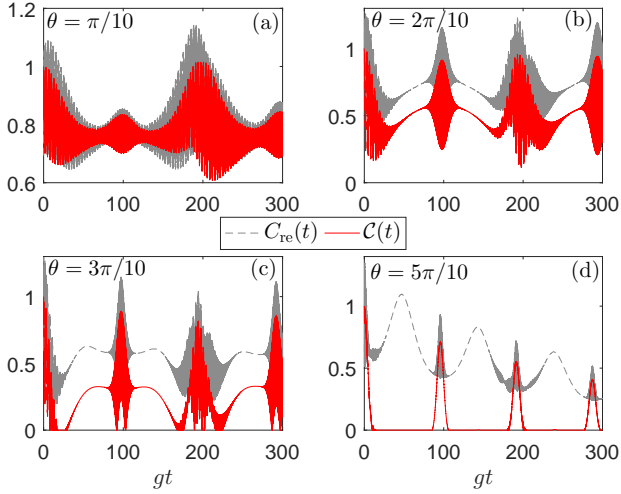}
\caption{Time evolution of the relative entropy of coherence $C_{\rm{re}}(t)$ (dashed gray curves) and the concurrence $\mathcal{C}(t)$ (red curves) for two qubits coupled to a common spin bath via the $XXX$-type coupling. The initial state is chosen as the maximally entangled state $|\varphi\rangle=\frac{1}{\sqrt{2}}(|\uparrow\downarrow\rangle+|\downarrow\uparrow\rangle)$ and we choose $g_1=g_2=g$, $\omega_1=\omega_2=\omega$, and $g'_1=g'_2=g'$. Parameters: $N=60$, $\omega/g=1$, $g'/g=1$, and $J/g=J'/g=0$.}
\label{Fig7}
\end{figure}
and focus on the case of $XXX$-type coupling. In Fig.~\ref{Fig7} we show the disentanglement dynamics of the two qubits starting with the maximally entangled state $|\varphi\rangle=(|\uparrow\downarrow\rangle+|\downarrow\uparrow\rangle)/\sqrt{2}$. Unlike the case of individual baths for which the profiles of the disentanglement dynamics behave similarly for different $\theta$'s, here the entanglement dynamics depends sensitively on $\theta$. For $\theta=\pi/10$, the bath initial state $|\hat{\Omega}\rangle$ is close to the fully polarized state $|L\rangle$ so that the time evolution mainly takes place in the subspace spanned by the basis states in the last few lines of Eq.~(\ref{basis2q}), which results in a oscillatory behavior between 1 and 0.5. For $\theta=2\pi/10$, we observe collapse and revival behaviors for both the concurrence and the coherence with period $\approx L\pi$. As $\theta$ is increased further, sudden disappearance and revival of the concurrence occurs. We note that collapse and revival behaviors in the entanglement dynamics of two qubits coupled to a single bosonic mode was revealed in Ref.~\cite{Sola2006}.

\par
\begin{figure}
\includegraphics[width=.54\textwidth]{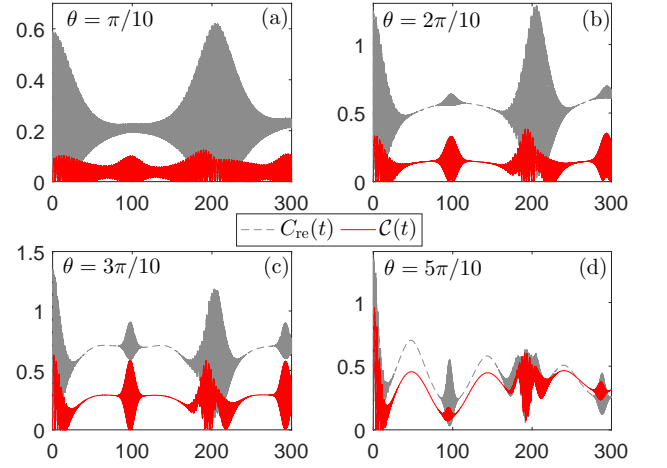}
\caption{The same as in Fig.~\ref{Fig7}, but for a separable two-qubit initial state $|\varphi\rangle=|\uparrow\uparrow\rangle$.}
\label{Fig8}
\end{figure}

 In the case of individual baths, two initially unentangled noninteracting qubits can never get entangled. In contrast, it is well known that two noninteracting qubits coupled to a common spin bath can become entangled even starting with an unentangled state. Figure~\ref{Fig8} shows the entanglement generation of two initially separable qubits prepared in the state $|\psi(0)\rangle=|\uparrow\uparrow\rangle$. We again observe the collapse and revival phenomena in the dynamics of both $C_{\mathrm{re}}(t)$ and $\mathcal{C}(t)$ for large values of $\theta$. Moreover, $\mathcal{C}(t)$ can keep a finite steady value for a relatively long time period of time in the duration of the collapse [Fig.~\ref{Fig8}(c) and (d)]. This provides a possible scenario to create steady entanglement between the two qubits via their couplings to the common spin bath.

\section{Intrabath entanglement dynamics}\label{SecIV}
\par In this section, we will study the dynamics of pairwise entanglement between two bath spins in the single- and two-qubit CSMs. To discuss the relation between the decoherence properties of the central system and the intrabath entanglement, we assume that the spin bath is initialized in an entangled state. To be specific, we focus on two types of entangled bath states, i.e., an equally weighted linear superposition of the Dicke states and the so-called $W$-class states~\cite{Milburn2005}.
\subsection{Equally weighted superposition state}
\par We first consider the following equally weighted superposition of the $N+1$ Dicke states
\begin{eqnarray}\label{eWphiB}
|\varphi_B\rangle=\frac{1}{\sqrt{N+1}}\sum^N_{n=0}|n\rangle_D,
\end{eqnarray}
as a bath initial state. This results in identical matrix elements of $\rho^{(1)}_B(0)$ (take the single-qubit CSM as an example), $[\rho^{(1)}_B(0)]_{nn'}=1/(N+1),~\forall n,n'$. It is easy to show that the corresponding reduced density matrix of $\vec{T}_1$ and $\vec{T}_2$ reads
\begin{eqnarray}
\chi^{(1)}(0)=\left(
                \begin{array}{cccc}
                  \frac{1}{3} & q_1 & q_1 & q_2 \\
                  q_1 & \frac{1}{6} & \frac{1}{6}  & q_1 \\
                  q_1 & \frac{1}{6}  & \frac{1}{6} & q_1 \\
                  q_2 & q_1 & q_1 & \frac{1}{3} \\
                \end{array}
              \right),
\end{eqnarray}
where $q_1=\frac{1}{N+1}\sum^{N-2}_{m=0}\frac{C^m_{N-2}}{\sqrt{C^{m+1}_NC^m_N}}$ and $q_2=\frac{1}{N+1}\sum^{N-2}_{m=0}\frac{C^m_{N-2}}{\sqrt{C^{m+2}_NC^m_N}}$. The two bath spins $\vec{T}_1$ and $\vec{T}_2$ are thus initially entangled. To track the evolution of the intrabath entanglement measured by the concurrence $\mathcal{C}^{(1)}_{12}(t)$ between $\vec{T}_1$ and $\vec{T}_2$, we define the reduced concurrence
\begin{eqnarray}
\tilde{\mathcal{C}}^{(1)}_{12}(t)\equiv \mathcal{C}^{(1)}_{12}(t)/\mathcal{C}^{(1)}_{12}(0),
\end{eqnarray}
which satisfies $\tilde{\mathcal{C}}^{(1)}_{12}(0)=1$.
\begin{figure}
\includegraphics[width=.54\textwidth]{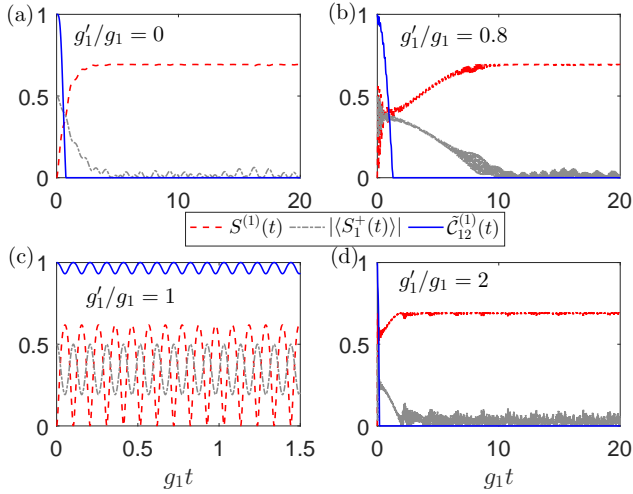}
\caption{Dynamics of the central-qubit von Neumann entropy $S^{(1)}(t)$ measuring the qubit-bath entanglement (red), the central-qubit coherence $|\langle S^+_1(t)\rangle|$ (gray), and the reduced intrabath entanglement $\tilde{\mathcal{C}}^{(1)}_{12}(t)$ (blue) in the single-qubit CSM. The initial states of the central qubit and the spin bath are chosen as $|\varphi^{(1)}\rangle=(|\uparrow\rangle+|\downarrow\rangle)/\sqrt{2}$ and $|\varphi_B\rangle$ given by Eq.~(\ref{eWphiB}), respectively. Other parameters: $N=60$, $\omega_1/g_1=0$. }
\label{Fig9}
\end{figure}
\par In Fig.~\ref{Fig9} we plot the dynamics of the von Neumann entropy of the central qubit, $S^{(1)}(t)=-\mathrm{Tr}[\rho^{(1)}(t)\ln \rho^{(1)}(t)]$, the coherence of the central qubit, $|\langle S^+_1(t)\rangle|$, and the reduced concurrence between two individual bath spins, $\tilde{\mathcal{C}}^{(1)}_{12}(t)$. The initial state is chosen as $|\psi^{(1)}_0\rangle=|\varphi^{(1)}\rangle|\varphi_B\rangle$, where $|\varphi^{(1)}\rangle=(|\uparrow\rangle+|\downarrow\rangle)/\sqrt{2}$ and $|\varphi_B\rangle$ is given by Eq.~(\ref{eWphiB}). It can be seen that for $g'_1/g_1=0$, $0.8$, and $2$, the intrabath entanglement drops abruptly to nearly zero in short times, which is accompanied by the corresponding decrease in the single-qubit coherence. This means that the decoherence properties of the central qubit is related to the disentanglement of the spin bath. As expected, a nearly steady system-bath entanglement is established right after the central spin lose its coherence, indicating that the initial entanglement within the bath is converted to the entanglement between the central spin and the bath. For $g'_1/g_1=1$ and $\omega_1=0$ we observe perfect oscillations for all the three quantities with Rabi frequency $A_n=g_1(2L+1)$. We also observe that the crest of $\tilde{\mathcal{C}}^{(1)}_{12}(t)$ is consistent with the crest (trough) of $|\langle S^+_1(t)\rangle|$ [$S^{(1)}(t)$], which again indicates a positive (negative) correlation between the intrabath entanglement and the central-spin coherence (the system-bath entanglement).

\par
\begin{figure}
\includegraphics[width=.54\textwidth]{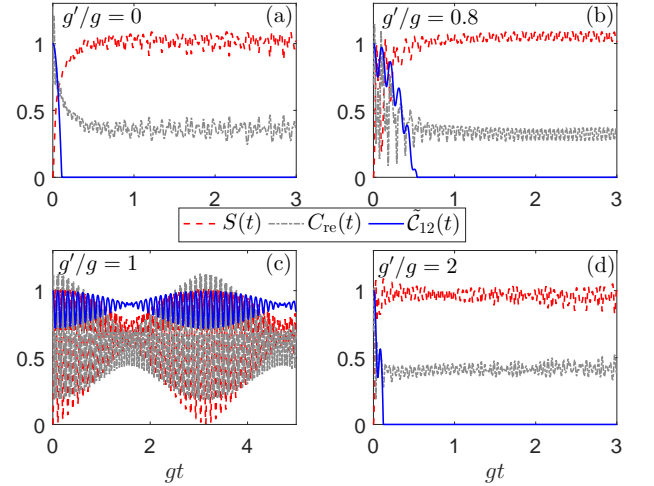}
\caption{Dynamics of the central-two-qubit von Neumann entropy $S(t)$ (red), the relative entropy of coherence of the two-qubit system $C_{\mathrm{re}}(t)$ (gray), and the reduced intrabath entanglement $\tilde{\mathcal{C}}_{12}(t)$ (blue) in the two-qubit CSM. The initial states of the two-qubit system and the spin bath are chosen as $|\varphi\rangle=(|\uparrow\downarrow\rangle+|\downarrow\uparrow\rangle)/\sqrt{2}$ and $|\varphi_B\rangle$ given by Eq.~(\ref{eWphiB}), respectively. Other parameters: $N=60$, $g_1=g_2=g$, $\omega_1/g=\omega_2/g=0$. }
\label{Fig9b}
\end{figure}

 Figure~\ref{Fig9b} shows the corresponding results for the two-qubit CSM, where the time evolutions of the two-qubit von Neumann entropy $S(t)$ and relative entropy of coherence $C_{\mathrm{re}}(t)$, as well as the reduced concurrence of two bath spins, $\tilde{\mathcal{C}}_{12}(t)\equiv \mathcal{C}_{12}(t)/\mathcal{C}_{12}(0)$, are plotted. The initial state of the two-qubit is chosen as $|\varphi\rangle=(|\uparrow\downarrow\rangle+|\downarrow\uparrow\rangle)/\sqrt{2}$. Despite the relative entropy of coherence approaches a finite steady value with minor oscillators, the relation among these three quantities is similar to that in the case of the single-qubit CSM. The numerical results indicate that the intrabath entanglement stored in the equally weighted superposition state given by Eq.~\ref{eWphiB} is generally fragile against the interaction to the central qubit.

\subsection{$W$-class states}
\par The second type of entangled state of the spin bath is the $W$-class state of the form
\begin{eqnarray}\label{phiBW}
|\varphi_B\rangle=\gamma_{N-1}|N-1\rangle_D+\gamma_N|N\rangle_D,
\end{eqnarray}
where $|\gamma_{N-1}|^2+|\gamma_N|^2=1$. The reduced density matrix of $\vec{T}_1$ and $\vec{T}_2$ corresponding to $|\varphi_B\rangle$ is
\begin{eqnarray}
\chi^{(1)}(0)=\left(
                \begin{array}{cccc}
                  0 & 0 & 0 & 0 \\
                  0 & \frac{|\gamma_{N-1}|^2}{N} & \frac{|\gamma_{N-1}|^2}{N}  & \frac{\gamma_{N-1}\gamma^*_N}{\sqrt{N}} \\
                  0 &\frac{|\gamma_{N-1}|^2}{N}  & \frac{|\gamma_{N-1}|^2}{N} & \frac{\gamma_{N-1}\gamma^*_N}{\sqrt{N}} \\
                  0 & \frac{\gamma^*_{N-1}\gamma_N}{\sqrt{N}} & \frac{\gamma^*_{N-1}\gamma_N}{\sqrt{N}} & 1-\frac{2|\gamma_{N-1}|^2}{N} \\
                \end{array}
              \right),
\end{eqnarray}
which gives the initial concurrence $\mathcal{C}_{12}(0)=\frac{2|\gamma_{N-1}|^2}{N}$.
\begin{figure}
\includegraphics[width=.54\textwidth]{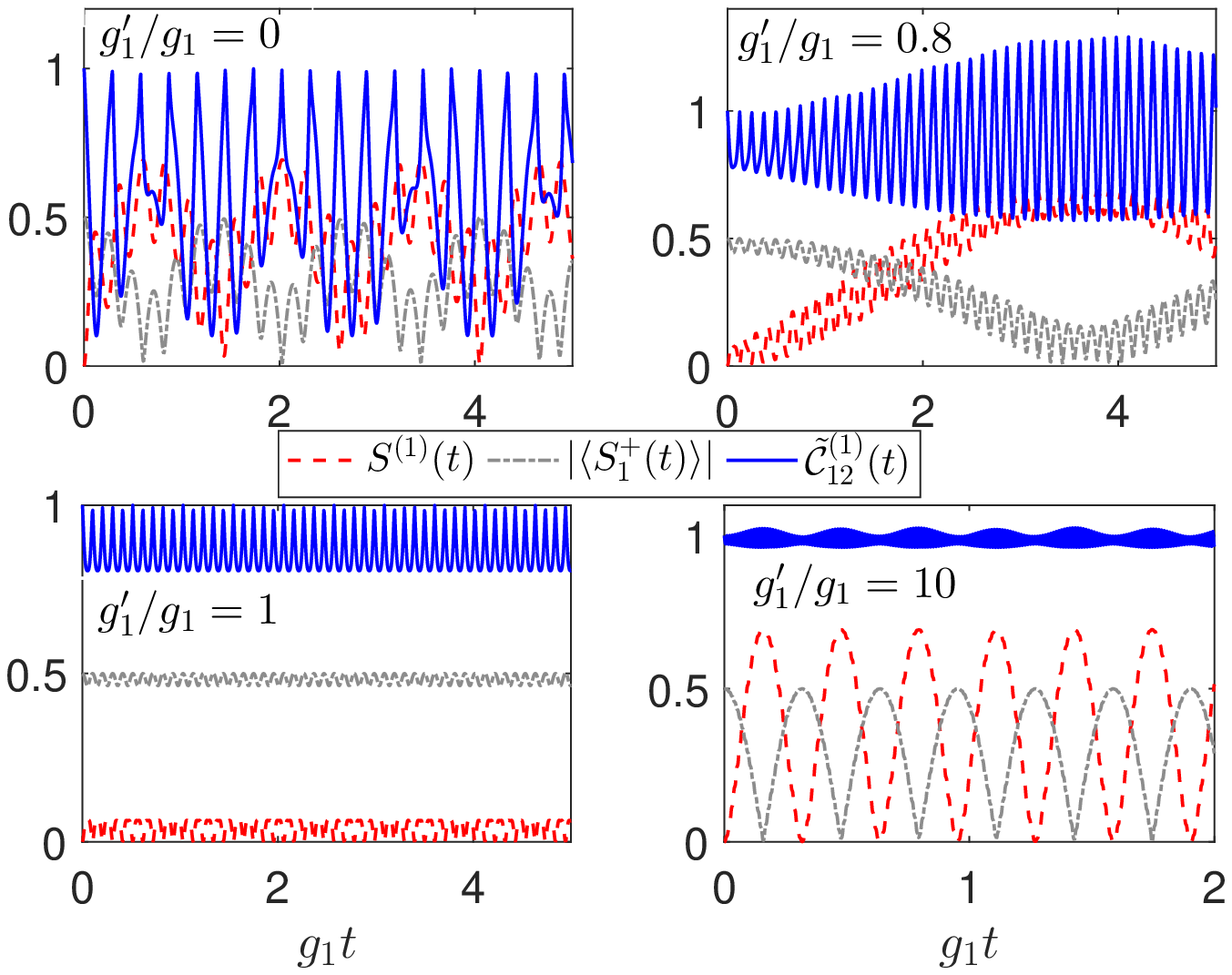}
\caption{The same as in Fig.~\ref{Fig9}, but for a $W$-class state $|\varphi_B\rangle=(|N-1\rangle_D+|N\rangle_D)/\sqrt{2}$.}
\label{Fig10}
\end{figure}
\begin{figure}
\includegraphics[width=.54\textwidth]{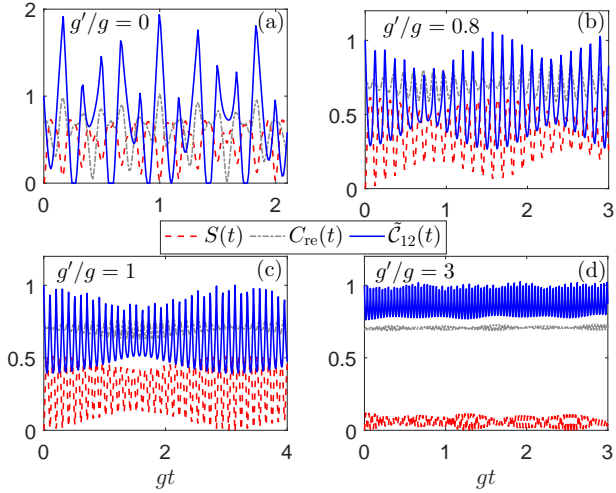}
\caption{The same as in Fig.~\ref{Fig9b}, but for a $W$-class state $|\varphi_B\rangle=(|N-1\rangle_D+|N\rangle_D)/\sqrt{2}$.}
\label{Fig10b}
\end{figure}
\par Figure~\ref{Fig10} shows the evolution of $S^{(1)}(t)$,  $|\langle S^+_1(t)\rangle|$, and $\tilde{\mathcal{C}}^{(1)}_{12}(t)$ for the initial state $|\psi^{(1)}_0\rangle=|\varphi^{(1)}\rangle|\varphi_B\rangle$, where $|\varphi^{(1)}\rangle=(|\uparrow\rangle+|\downarrow\rangle)/\sqrt{2}$ and $|\varphi_B\rangle$ is given by Eq.~(\ref{phiBW}) with $\gamma_{N-1}=\gamma_{N}=1/\sqrt{2}$. Unlike in the case of the equally weighted bath initial state, all the three quantities oscillates in a nearly regular way and no complete disentanglement of the bath occurs. For $g'_1/g_1\gg 1$ [Fig.~\ref{Fig10}(d)], the $XXZ$ CSM reduces to the Ising CSM used by Zurek to discuss spin environment-induced superselection rules~\cite{Zurek} and decoherence~\cite{Zurek2005}, the intrabath concurrence $ \mathcal{C}^{(1)}_{12}(t)$ thus nearly keeps its initial value during the time evolution due to the dominance of the non-spin-flipping Ising term~\cite{Milburn2005}. Figure~\ref{Fig10b} shows the corresponding results for the two-qubit CSM and similar behaviors are observed. We still see a negative (positive) correlation between the intrabath entanglement and the system-bath entanglement (coherence of the central system). In contrast to the case of equally weighted bath initial state, the $W$-class states are robust in exchanging the intrabath entanglement and the system-bath entanglement.
\section{Conclusions}\label{SecV}
\label{sec-final}
\par In this work, we obtain benchmark exact quantum dynamics of the two-qubit $XXZ$ homogeneous central spin model with the spin bath prepared in linear superpositions of symmetric Dicke states. By working in the interaction picture with respect to the non-spin-flipping part of the Hamiltonian, we derive a set of equations of motion satisfied by the probability amplitudes of the time-evolved wave function of the whole system. For a single-qubit central spin problem, the equations of motion can be analytically solved to give closed-form expressions for the probability amplitudes, recovering the results in Ref.~\cite{Guan2019}. These analytical results are then used to study the time-evolution of the single-qubit polarization and purity dynamics. The observed collapse and revival phenomena are explained in a quantitative way following an analogical analysis in Ref.~\cite{PRL1990} on the Jaynes-Cummings model for as spin bath prepared in the spin coherent state.
\par We then study the disentanglement dynamics of two initially entangled noninteracting qubits interacting with individual baths. The dynamics of such a system can be directly determined by the single-qubit dynamics following the procedures presented in Ref.~\cite{nonmark}. We calculate the time-evolution of the concurrence and relative entropy of coherence of the two-qubit system and find that the two-qubit coherence behaves similarly to the single-qubit purity. Due to the non-Markovian nature of the spin bath, we observe entanglement sudden disappearance in a finite time and the subsequent revivals. We also study the entanglement generation process between the two initially unentangled qubits caused by the coupling to a common spin bath. For certain polarizations of the spin coherent states, we observe collapse and revival behaviors in the entanglement and coherence dynamics, which provides a possible setup to realized steady and finite two-qubit entanglement or coherence over long periods of time.
\par We finally investigate the entanglement dynamics of two individual bath spins for initially entangled baths. In particular, we choose the equally weighted superposition of the symmetric Dicke state and the $W$-class states for which the initial entanglement between any two bath spins is nonvanishing. We find that for the former case the intrabath entanglement is fragile with respect to the influence of the central system, i.e., the entanglement between any pair of bath spins decays to zero in a short time scale, accompanied by an increase in the system-bath entanglement and a decrease in the central-system coherence. For the $W$-class states, we observe a robust exchange between the intrabath entanglement and the system-entanglement.

\noindent{\bf Acknowledgments:}
This work was supported by the Natural Science Foundation of China (NSFC) under Grant No. 11705007, and partially by the Beijing Institute of Technology Research Fund Program for Young Scholars. W.-L. Y. acknowledge support from the start-up fund of Nanjing University of Aeronautics and Astronautics (Grant No. 1008-YAH20006).

\appendix
\section{On the spin coherent state}\label{AppA}
\par The spin coherent state of the $N$ spins-1/2 is defined as
\begin{eqnarray}\label{SCSdef}
|\hat{\Omega}\rangle&=&e^{-iL_z\phi}e^{-iL_y\theta}|\frac{N}{2},\frac{N}{2}\rangle\nonumber\\
&=&\prod^{N}_{j=1}e^{-iT^z_j\phi}e^{-iT^y_j\theta}|+\rangle_j\nonumber\\
&=&\prod^{N}_{j=1}\left[\cos\frac{\theta}{2} e^{-i\phi/2}|+\rangle_j+\sin\frac{\theta}{2} e^{i\phi/2}|-\rangle_j\right],
\end{eqnarray}
where $|+\rangle_j$ is the spin-up state for the $j$th bath spin. By expanding the right-hand side of Eq.~(\ref{SCSdef}), $|\hat{\Omega}\rangle$ can be rewritten as
\begin{eqnarray}\label{SCSdef1}
|\hat{\Omega}\rangle&=&\sum^N_{m=0}\left(\cos\frac{\theta}{2}\right)^m\left(\sin\frac{\theta}{2}\right)^{N-m}e^{i\phi(N-2m)/2}\nonumber\\
&&\sum_{n_1<n_2<\cdots<n_{N-m}}T^-_{n_1}\cdots T^-_{n_{N-m}}\left(\prod^N_{j=1}|+\rangle_j\right).\nonumber\\
\end{eqnarray}
Recall that the Dicke states can be expressed as
\begin{eqnarray}
&&|\frac{N}{2},m-\frac{N}{2}\rangle\nonumber\\
&=&\frac{1}{\sqrt{C^m_N}}\sum_{n_1<n_2<\cdots<n_{N-m}}T^-_{n_1}\cdots T^-_{n_{N-m}}\left(\prod^N_{j=1}|+\rangle_j\right),\nonumber\\
\end{eqnarray}
we thus have
\begin{eqnarray}\label{SCSdef2}
|\hat{\Omega}\rangle&=&\sum^N_{m=0}\left(\cos\frac{\theta}{2}\right)^m\left(\sin\frac{\theta}{2}\right)^{N-m}\nonumber\\
&&e^{i\phi(N-2m)/2}\sqrt{C^m_N}|\frac{N}{2},m-\frac{N}{2}\rangle.
\end{eqnarray}
\par Up to a global phase factor $e^{i\phi N/2}$, the spin coherent state $|\hat{\Omega}\rangle$ can finally be written as
\begin{eqnarray}\label{SCSdef3}
|\hat{\Omega}\rangle&=&\sum^N_{n=0}Q_n|\frac{N}{2},n-\frac{N}{2}\rangle,
\end{eqnarray}
where
\begin{eqnarray}
Q_n&\equiv&\frac{z^n}{(1+|z|^2)^{N/2}}\sqrt{C^n_N},
\end{eqnarray}
with $z=\cot\frac{\theta}{2}e^{-i\phi}$.
\section{Equations of motion for a single-qubit in the $M=m-\frac{1}{2}$ sector}\label{Singlem}
\par Consider a typical evolved state governed by $H^{(1)}_I(t)$ in the $M=m-\frac{1}{2}$ sector (let $|m\rangle=|\frac{N}{2},m\rangle$),
\begin{eqnarray}
|\chi^{(m)}(t)\rangle= C^{(m)}_{\uparrow}(t)|\uparrow\rangle|m-1\rangle+C^{(m)}_{\downarrow}(t)|\downarrow\rangle|m\rangle.
\end{eqnarray}
Applying the Schr\"odinger operator $H^{(1)}_I(t)$ to the above state, we get
\begin{eqnarray}
&&H^{(1)}_I(t)|\chi^{(m)}(t)\rangle\nonumber\\
&=& C^{(m)}_{\uparrow}(t)g_1e^{-i\omega_1t} e^{-i2g'_1mt}e^{ i g'_1 t}x_{l,m-1}|\downarrow\rangle|m\rangle\nonumber\\
&&+C^{(m)}_\downarrow(t)g_1x_{l,m-1}e^{- i g'_1 t}e^{ i2g'_1mt}e^{ i\omega_1t}  |\uparrow\rangle|m-1\rangle.\nonumber\\
\end{eqnarray}
From the Schr\"odinger equation $i\partial_t|\chi^{(m)}(t)\rangle=H^{(1)}_I(t)|\chi^{(m)}(t)\rangle$, we obtain the following equations of motion for the coefficients $C^{(m)}_{\uparrow/\downarrow}(t)$
\begin{eqnarray}
i\dot{C}^{(m)}_{\uparrow}(t)&=&C^{(m)}_-(t)g_1x^{(-)}_{l,m-1}e^{- i g'_1 t}e^{ i2g'_1mt}e^{ i\omega_1t},\nonumber\\
i\dot{C}^{(m)}_{\downarrow}(t)&=&C^{(m)}_{+}(t)g_1 x^{(-)}_{l,m-1}e^{-i\omega_1t} e^{-i2g'_1mt}e^{ i g'_1 t}.\nonumber\\
\end{eqnarray}
\section{Analytical solutions of Eq.~(\ref{FF})}\label{ODE}
\par We provide the analytical solution to the following two coupled first-order ordinary differential equations
\begin{eqnarray}
i\dot{x}_1&=&a e^{ibt}x_2,\nonumber\\
i\dot{x}_2&=&a e^{-ibt}x_1,
\end{eqnarray}
under the initial condition
\begin{eqnarray}
x_1(0)=X_1,~~x_2(0)=X_2,
\end{eqnarray}
where $a$ and $b$ are two real constants.
\par We first differentiate the first equation once to get
\begin{eqnarray}
\ddot{x}_1-ib\dot{x}_1+a^2x_1&=&0.
\end{eqnarray}
The characteristic equation for the last equation is
\begin{eqnarray}
\lambda^2-ib\lambda+a^2=0,
\end{eqnarray}
which gives two distinct imaginary roots
\begin{eqnarray}
\lambda_\pm=\frac{i}{2}(b\pm\sqrt{b^2+4a^2}) 
\end{eqnarray}
So the general solution is
\begin{eqnarray}\label{x1}
x_1=c_+ e^{\lambda_+t}+c_- e^{\lambda_-t}
\end{eqnarray}
with
\begin{eqnarray}\label{X1}
c_++c_-=X_1.
\end{eqnarray}
The other variable $x_2(t)$ can obtained as
\begin{eqnarray}\label{x2}
x_2&=&\frac{i}{a}e^{-ibt}\dot{x}_1\nonumber\\
&=&\frac{i}{a}e^{-ibt}(c_+ \lambda_+e^{\lambda_+t}+c_-\lambda_- e^{\lambda_-t})
\end{eqnarray}
so that
\begin{eqnarray}\label{X2}
X_2= \frac{i}{a} (c_+ \lambda_+ +c_-\lambda_-  )
\end{eqnarray}
Solving Eqs.~(\ref{X1}) and (\ref{X2}) gives
\begin{eqnarray}
c_-&=&-i\frac{\lambda_+}{A}X_1+\frac{a}{A}X_2,\nonumber\\
c_+&=&\frac{ i\lambda_-}{A}X_1- \frac{a}{A}X_2,
\end{eqnarray}
where $A\equiv\sqrt{b^2+4a^2}$. By inserting the coefficients $c_\pm$ into Eqs.~(\ref{x1}) and (\ref{x2}), we finally get the solution
\begin{eqnarray}
x_1&=&e^{\frac{i}{2}bt}\left[ X_1  \cos\frac{At}{2}-i\left(\frac{b}{A}X_1+\frac{2a}{A}X_2\right)\sin\frac{At}{2}\right],\nonumber\\
x_2&=& e^{-\frac{i}{2}bt}\left[   X_2 \cos\frac{At}{2}+ i \left(\frac{-2a }{A}X_1+\frac{  b}{A}X_2\right)\sin\frac{At}{2} \right].\nonumber\\
\end{eqnarray}
\section{Explicit forms of the coefficients $\{Y_{abcd}(t)\}$ in Eq.~(\ref{rhotrho0})}\label{AppD}
Direct calculations based on Eqs.~(\ref{FFsol}) and (\ref{FFsoltil}) give the following explicit forms of $Y_{abcd}(t)$:
\begin{eqnarray}\label{YY1}
Y_{\uparrow\uparrow\uparrow\uparrow}&=&\sum^{N}_{n=0}|Q_{n}|^2 |W_{n+1}|^2,~Y_{\uparrow\uparrow\downarrow\downarrow}=\sum^{N}_{n=1}|Q_{n}|^2 |V_{n}|^2,\nonumber\\
Y_{\uparrow\uparrow\uparrow\downarrow}&=&Y^*_{\uparrow\uparrow\downarrow\uparrow}=\sum^{N}_{n=1}Q_{n-1}Q^*_{n}W_{n}V^*_{n},
\end{eqnarray}
\begin{eqnarray}\label{YY2}
Y_{\downarrow\downarrow\uparrow\uparrow}&=&\sum^N_{n=1} |Q_{n-1}|^2|V_n|^2,~Y_{\downarrow\downarrow\downarrow\downarrow}=\sum^N_{n=0} |Q_{n}|^2|W_n|^2,\nonumber\\
Y_{\downarrow\downarrow\uparrow\downarrow}&=&Y^*_{\downarrow\downarrow\downarrow\uparrow}=\sum^N_{n=1}Q_{n-1}Q^*_n W_n V_n ,
\end{eqnarray}
and
\begin{eqnarray}\label{YY3}
Y_{\uparrow\downarrow\uparrow\uparrow}&=&Y^*_{\downarrow\uparrow\uparrow\uparrow}=\sum^N_{n=1} Q_n Q^*_{n-1}W_{n+1}V^*_n,\nonumber\\
Y_{\uparrow\downarrow\downarrow\downarrow}&=&Y^*_{\downarrow\uparrow\downarrow\downarrow}=\sum^{N}_{n=1} Q^*_{n-1}Q_{n}W_{n-1}V_{n},\nonumber\\
Y_{\uparrow\downarrow\uparrow\downarrow}&=&Y^*_{\downarrow\uparrow\downarrow\uparrow}=\sum^N_{n=0}|Q_n|^2W_nW_{n+1},\nonumber\\
Y_{\uparrow\downarrow\downarrow\uparrow}&=&Y^*_{\downarrow\uparrow\uparrow\downarrow}=\sum^{N-1}_{n=1}Q^*_{n-1}Q_{n+1}V_{n}^*V_{n+1}.
\end{eqnarray}
\section{Derivation of the matrix representation of $T^+_1T^-_1T^+_2T^-_2$ etc. in the Dicke basis}\label{AppE}
\par To derive the matrix representation of, e.g., $T^+_1T^-_1T^+_2T^-_2$, in the Dicke basis, we need to consider the action of $T^+_1T^-_1T^+_2T^-_2$ on the Dicke state $|m\rangle_D$, which can be rewritten as
\begin{eqnarray}
&&|m\rangle_D=\frac{|+\rangle_1|+\rangle_2}{\sqrt{C^m_N}}\sum_{j_1<\cdots<j_m (\neq 1,2)}T^-_{j_1}\cdots T^-_{j_m}|\prod^N_{l(\neq 1,2)}|+\rangle_l\nonumber\\
&&+\frac{|+\rangle_1|-\rangle_2}{\sqrt{C^m_N}}\sum_{j_1<\cdots<j_{m-1} (\neq 1,2)}T^-_{j_1}\cdots T^-_{j_{m-1}}|\prod^N_{l(\neq 1,2)}|+\rangle_l\nonumber\\
&&+\frac{|-\rangle_1|+\rangle_2}{\sqrt{C^m_N}}\sum_{j_1<\cdots<j_{m-1} (\neq 1,2)}T^-_{j_1}\cdots T^-_{j_{m-1}}|\prod^N_{l(\neq 1,2)}|+\rangle_l\nonumber\\
&&+\frac{|-\rangle_1|-\rangle_2}{\sqrt{C^m_N}}\sum_{j_1<\cdots<j_{m-2} (\neq 1,2)}T^-_{j_1}\cdots T^-_{j_{m-2}}|\prod^N_{l(\neq 1,2)}|+\rangle_l.\nonumber\\
\end{eqnarray}
The number of terms in the sums of the above four lines are $C^{m}_{N-2}$, $C^{m-1}_{N-2}$, $C^{m-1}_{N-2}$, and $C^{m-2}_{N-2}$, respectively. This results in $~_D\langle m|T^+_1T^-_1T^+_2T^-_2|m\rangle_D=C^m_{N-2}/C^m_N$. The matrix representations of $T^+_1T^-_1T^-_2T^+_2$, $T^-_1T^+_1T^+_2T^-_2$, $T^-_1T^+_1T^-_2T^+_2$, and $T^-_1T^+_2$ can be similarly obtained and are all diagonal. Actually, we find that the matrix representations of  $T^+_1T^-_1T^-_2T^+_2$, $T^-_1T^+_1T^+_2T^-_2$, and $T^-_1T^+_2$ are identical in the Dicke basis.
\par The four triple operators $T^+_1T^-_1T^-_2$, $T^-_1T^+_2T^-_2$, $T^-_1T^-_2T^+_2$, and $T^-_1T^+_1T^-_2$ do not conserve $L_z$ and result in states having $m+1$ down spins when they act on $|m\rangle_D$. So the only nonvanishing matrix elements are
\begin{eqnarray}
~_D\langle m+1|T^+_1T^-_1T^-_2|m\rangle_D&=&~_D\langle m+1|T^-_1T^+_2T^-_2|m\rangle_D\nonumber\\
&=&C^m_{N-2}/\sqrt{C^{m+1}_N C^m_N},\nonumber\\
~_D\langle m+1|T^-_1T^-_2T^+_2|m\rangle_D&=&~_D\langle m+1|T^-_1 T^+_1T^-_2|m\rangle_D\nonumber\\
&=&C^{m-1}_{N-2}\sqrt{C^{m+1}_N C^m_N}.
\end{eqnarray}
\par Finally, the operator $T^-_1T^-_2$ does not conserve $L_z$ either. When acting on $|m\rangle_D$, it results in a state having $m+2$ down spins, resulting in the nonvanishing matrix element
\begin{eqnarray}
~_D\langle m+2|T^-_1T^-_2|m\rangle_D&=&C^{m}_{N-2}/\sqrt{C^{m+2}_N C^m_N}.
\end{eqnarray}


\begin{thebibliography}{99}
\bibitem{spinbath2000} N. V. Prokof’ev and P. C. E. Stamp, Rep. Prog. Phys. \textbf{63}, 669 (2000).
\bibitem{Zurek2005} F. M. Cucchietti, J. P. Paz, and W. H. Zurek, Phys. Rev. A \textbf{72}, 052113 (2005).
\bibitem{Milburn2005}  C. M. Dawson, A. P. Hines, R. H. McKenzie, and G. J. Milburn, Phys. Rev. A \textbf{71}, 052321 (2005).
\bibitem{Sun2006} H. T. Quan, Z. Song, X. F. Liu, P. Zanardi, and C. P. Sun, Phys. Rev. Lett. \textbf{96}, 140604 (2006).
\bibitem{Petruccione2008} Y. Hamdouni and F. Petruccione, Phys. Rev. B \textbf{76}, 174306 (2007).
\bibitem{Takahashi2008} S. Takahashi, R. Hanson, J. van Tol, M. S. Sherwin, and D. D. Awschalom, Phys. Rev. Lett. \textbf{101}, 047601 (2008).
\bibitem{Sarma2008} W. M. Witzel and S. Das Sarma, Phys. Rev. B \textbf{77}, 165319 (2008).
\bibitem{Chen2008} C. Y. Lai, J. T. Hung, C. Y. Mou, and P. Chen, Phys. Rev. B \textbf{77}, 205419 (2008).
\bibitem{RBLiu2011} N. Zhao, Z.-Y. Wang, and R.-B. Liu, Phys. Rev. Lett. \textbf{106}, 217205 (2011).
\bibitem{F-S2013} A. Faribault and D. Schuricht, Phys. Rev. B \textbf{88}, 085323 (2013).
\bibitem{Wang2013} Z.-H. Wang and S. Takahashi, Phys. Rev. B \textbf{87}, 115122 (2013).
\bibitem{CRcat} S. Dooley, F. McCrossan, D. Harland, M. J. Everitt, and T. P. Spiller, Phys. Rev. A \textbf{87}, 052323 (2013).
\bibitem{PRA2014} N. Wu, A. Nanduri, and H. Rabitz, Phys. Rev. A \textbf{89}, 062105 (2014).
\bibitem{PRB2016} N. Wu, N. Fr\"ohling, X. Xing, J. Hackmann, A. Nanduri, F.B. Anders, and H. Rabitz, Phys. Rev. B \textbf{93}, 035430 (2016).
\bibitem{Cywski2018} D. Kwiatkowski and {\L}. Cywi\'nski, Phys. Rev. B \textbf{98}, 155202 (2018).
\bibitem{Yang2020} P. Lu, H.-L. Shi, L. Cao, X.-H. Wang, T. Yang, J. Cao, and W.-L. Yang, Phys. Rev. B \textbf{101}, 184307 (2020).
\bibitem{Loss2002} A. V. Khaetskii, D. Loss, and L. Glazman, Phys. Rev. Lett. \textbf{88}, 186802 (2002).
\bibitem{Loss2003} A. V. Khaetskii, D. Loss, and L. Glazman, Phys. Rev. B \textbf{67}, 195329 (2003).
\bibitem{Bose2004} A. Hutton and S. Bose, Phys. Rev. A \textbf{69}, 042312 (2004).
\bibitem{Messina2006} A. Napoli, F. Palumbo, and A. Messina, Journal of Physics: Conference Series \textbf{36}, 154 (2006).
\bibitem{Zhu2007} X. Z. Yuan, H. S. Goan, and K. D. Zhu, Phys. Rev. B \textbf{75}, 045331 (2007).
\bibitem{Cirac2008} H. Christ, J. I. Cirac, and G. Giedke, Phys. Rev. B \textbf{78}, 125314 (2008).
\bibitem{Liu2017} G.-Q. Liu, \emph{et al}., Phys. Rev. Lett. \textbf{118}, 150504 (2017).
\bibitem{Duan2018} F. Wang, \emph{et al}., Phys. Rev. B \textbf{98}, 064306 (2018).
\bibitem{Du2019} L. Dong, H. Liang, C.-K. Duan, Y. Wang, Z. Li, X. Rong, J. Du, Phys. Rev. A \textbf{99}, 013426 (2019).
\bibitem{Olaya2008} A. Olaya-Castro, C. F. Lee, F. F. Olsen, and N. F. Johnson, Phys. Rev. B \textbf{78}, 085115 (2008).
\bibitem{Sinayskiy2012} I. Sinayskiy, A. Marais, F. Petruccione, and A. Ekert, Phys. Rev. Lett. \textbf{108}, 020602 (2012).
\bibitem{Wu2013} N. Wu and Y. Zhao, J. Chem. Phys. \textbf{139}, 054118 (2013).
\bibitem{Petruccione2004} H.-P. Breuer, D. Burgarth, and F. Petruccione, Phys. Rev. B \textbf{70}, 045323 (2004).
\bibitem{Petruccione2006} Y. Hamdouni, M. Fannes, and F. Petruccione, Phys. Rev. B \textbf{73}, 245323 (2006).
\bibitem{Breuer2007} J. Fischer and H.-P. Breuer, Phys. Rev. A \textbf{76}, 052119 (2007).
\bibitem{Petruccione2007} H.-P. Breuer and F. Petruccione, Phys. Rev. E \textbf{76}, 016701 (2007).
\bibitem{Messina2008} E. Ferraro, H.-P. Breuer, A. Napoli, M. A. Jivulescu, and A. Messina, Phys. Rev. B \textbf{78}, 064309 (2008).
\bibitem{Petruccione2012} V. Semin, I. Sinayskiy, and F. Petruccione, Phys. Rev. A \textbf{86}, 062114 (2012).
\bibitem{Petruccione2014} V. Semin, I. Sinayskiy, and F. Petruccione, Phys. Rev. A \textbf{89}, 012107 (2014).
\bibitem{Gaudin1976} M. Gaudin, J. Phys. France \textbf{37}, 1087 (1976).
\bibitem{Kiss2001} D. Garajeu and A. Kiss, J. Math. Phys. \textbf{42}, 3497 (2001).
\bibitem{Ortiz2005} G. Ortiz, R. Somma, J. Dukelsky, and S. Rombouts, Nucl. Phys. B \textbf{707}, 421 (2005).
\bibitem{Bortz2007} M. Bortz and J. Stolze, Phys. Rev. B \textbf{76}, 014304 (2007).
\bibitem{Erbe2010} B. Erbe and J. Schliemann, Phys. Rev. Lett. \textbf{105}, 177602 (2010).
\bibitem{Claeys2015} P. W. Claeys, S. De Baerdemacker, M. Van Raemdonck, and D. Van Neck, Phys. Rev. B \textbf{91}, 155102 (2015).
\bibitem{Wu2018} N. Wu, Physica A \textbf{501}, 308 (2018).
\bibitem{Guan2018} R. I. Nepomechie and X.-W. Guan, J. Stat. Mech. (2018) 103104.
\bibitem{Skrypnyk2019} T. Skrypnyk, Nucl. Phys. B \textbf{941}, 225 (2019).
\bibitem{Guan2019} W.-B. He, S. Chesi, H.-Q. Lin, and X.-W. Guan, Phys. Rev. B \textbf{99}, 174308 (2019).
\bibitem{PRB2020} N. Wu, X.-W. Guan, and J. Links, Phys. Rev. B \textbf{101}, 155145 (2020).
\bibitem{Vill2020} T. Villazon, A. Chandran, and P. W. Claeys, arXiv:2001.10008.
\bibitem{Lidar2010} N. Arshed, A. H. Toor, and D. A. Lidar, Phys. Rev. A \textbf{81}, 062353 (2010).
\bibitem{Braun2002} D. Braun, Phys. Rev. Lett. \textbf{89}, 277901 (2002).
\bibitem{Piani2003} F. Benatti, R. Floreanini, and M. Piani, Phys. Rev. Lett. \textbf{91}, 070402 (2003).
\bibitem{erbly}  T. Yu and J. H. Eberly, Phys. Rev. Lett. \textbf{93}, 140404 (2004).
\bibitem{Kim2006} S. Oh and J. Kim, Phys. Rev. A \textbf{73}, 062306 (2006).
\bibitem{Science2007} M. P. Almeida, F. de Melo, M. Hor-Meyll, A. Salles, S. P. Walborn, P. H. Souto Ribeiro, L. Davidovich, Science \textbf{316}, 579 (2007).
\bibitem{nonmark}  B. Bellomo, R. Lo Franco, and G. Compagno, Phys. Rev. Lett. \textbf{99}, 160502 (2007).
\bibitem{Privman2007} D. Solenov, D. Tolkunov, and V. Privman, Phys. Rev. B \textbf{75}, 035134 (2007).
\bibitem{Dajka2008} J. Dajka, M. Mierzejewski, and J. {\L}uczka, Phys. Rev. A \textbf{77}, 042316 (2008).
\bibitem{Plastina2008} S. Maniscalco, F. Francica, R. L. Zaffino, N. Lo Gullo, and F. Plastina, Phys. Rev. Lett. \textbf{100}, 090503 (2008).
\bibitem{Aguado2008} L. D. Contreras-Pulido and R. Aguado, Phys. Rev. B \textbf{77}, 155420 (2008).
\bibitem{Nazir2009} D. P. S. McCutcheon, A. Nazir, S. Bose, and A. J. Fisher, Phys. Rev. A \textbf{80}, 022337 (2009).
\bibitem{Tanimura2010}  A. G. Dijkstra and Y. Tanimura, Phys. Rev. Lett. \textbf{104}, 250401 (2010).
\bibitem{Yu2011} X. Zhao, J. Jing, B. Corn, and T. Yu, Phys. Rev. A \textbf{84}, 032101 (2011).
\bibitem{Ma2012} J. Ma, Z. Sun, X. Wang, and F. Nori, Phys. Rev. A \textbf{85}, 062323 (2012).
\bibitem{Bellomo2013}  R. Lo Franco, B. Bellomo, S. Maniscalco, and G. Compagno, Int. J. Mod. Phys. B \textbf{27}, 1345053 (2013).
\bibitem{XBWang2013} N. Qiu and X.-B. Wang, Phys. Rev. A \textbf{88}, 062332 (2013).
\bibitem{Makri2013} M. M. Sahrapour and N. Makri, J. Chem. Phys. \textbf{138}, 114109 (2013).
\bibitem{Kast2014} D. Kast and J. Ankerhold, Phys. Rev. B \textbf{90}, 100301(R) (2014).
\bibitem{Yu2019} J. She, J. Hu, and H. Yu, Phys. Rev. D \textbf{99}, 105009 (2019).
\bibitem{Lidar2007} H. Krovi, O. Oreshkov, M. Ryazanov, and D. A. Lidar, Phys. Rev. A \textbf{76}, 052117 (2007).
\bibitem{Breuer2008} E. Ferraro, H.-P. Breuer, A. Napoli, M. A. Jivulescu, and A. Messina, Phys. Rev. B \textbf{78}, 064309 (2008).
\bibitem{Lorenzo2013} S. Lorenzo, F. Plastina, and M. Paternostro, Phys. Rev. A \textbf{87}, 022317 (2013).
\bibitem{Dicke}  R. H. Dicke, Phys. Rev. \textbf{93}, 99 (1954).
\bibitem{PRA1972}  F. T. Arecchi, E. Courtens, R. Gilmore, and H. Thomas, Phys. Rev. A \textbf{6}, 2211 (1972).
\bibitem{Wstate} W. Dur, G. Vidal and J.I. Cirac, Phys. Rev. A \textbf{62}, 062314 (2000).
\bibitem{catstate} D. D. Bhaktavatsala Rao, N. Bar-Gill, and G. Kurizki, Phys. Rev. Lett. \textbf{106}, 010404 (2011).
\bibitem{QB} D. Ferraro, M. Campisi, G. M. Andolina, V. Pellegrini, and M. Polini, Phys. Rev. Lett. \textbf{120}, 117702 (2018).
\bibitem{PRL1990}  J. Gea-Banacloche, Phys. Rev. Lett. \textbf{65}, 3385 (1990).
\bibitem{Wootters}  W. K. Wootters, Phys. Rev. Lett. \textbf{80}, 2245 (1998).
\bibitem{Plenio}  T. Baumgratz, M. Cramer, and M. B. Plenio, Phys. Rev. Lett. \textbf{113}, 140401 (2014).
\bibitem{Bose2009}  H. Wichterich and S. Bose, Phys. Rev. A \textbf{79}, 060302(R) (2009).
\bibitem{Wu2010}  Z. Chang and N. Wu, Phys. Rev. A \textbf{81}, 022312 (2010).
\bibitem{Sola2006}  V. S. Malinovsky and I. R. Sola, Phys. Rev. Lett. \textbf{96}, 050502 (2006).
\bibitem{Zurek} W. H. Zurek, Phys. Rev. D \textbf{26}, 1862 (1982).
\bibitem{nonMark2004}  H.-P. Breuer, D. Burgarth, and F. Petruccione, Phys. Rev. B \textbf{70}, 045323 (2004).



\end{thebibliography}
\end{document}